\documentclass{article}
\usepackage{amsthm,amsmath,amsfonts,amssymb,graphicx,etoolbox}
\usepackage[authoryear]{natbib}
\setcitestyle{round}
\usepackage{indentfirst}
\usepackage[colorlinks=true,citecolor=blue,pdfpagemode=UseNone,pdfstartview=FitH]{hyperref}

\allowdisplaybreaks[4]

\title{A Conversation with A. Philip Dawid}
\author{Vladimir Vovk and Glenn Shafer}

\begin{document}
\maketitle
\begin{abstract}
  \noindent
  Beginning in the 1970s, Alexander Philip Dawid has been a leading contributor to the foundations of statistics
  and especially to the development and application of Bayesian statistics.
  He is also known for his work on causality,
  especially his notation for conditional independence and his critique of the overuse of counterfactuals,
  and for his contributions to forensic statistics.

  Dawid was born in Lancashire, England, on February 1, 1946.
  His family moved to London soon afterwards, and he attended the City of London School from 1956 to 1963.
  He studied mathematics at Cambridge, earning a BA (Bachelor of Arts) degree in 1966.
  After earning a Diploma in Mathematical Statistics in the academic year 1966--1967,
  he studied for a PhD at Imperial, then at UCL,
  where he became a Lecturer in Statistics in 1969.
  In 1978, he left UCL for a position as Professor of Statistics
  in the Department of Mathematics, The City University, London,
  where he served as Head of Statistics Section and Director of the Statistical Laboratory.
  He returned to the Department of Statistics at UCL in 1981,
  serving as Head of Department from 1983 to 1993.
  He moved to the University of Cambridge in 2007,
  becoming Professor of Statistics and Fellow of Darwin College.
  He has continued his work in mathematical statistics after retiring from Cambridge in 2013
  and was elected Fellow of the Royal Society in 2018.

  \medskip
  \noindent
  The most up-to-date version of this conversation
  is at \url{http://probabilityandfinance.com} (Working Paper 65).
\end{abstract}

\noindent
Both Vladimir Vovk and Glenn Shafer have known Philip Dawid personally for decades,
and his work has served as an inspiration and starting point
for much of their own work.
This conversation took place remotely in August and September of 2022.
Its definitive, much shorter, record is to appear in \emph{Statistical Science}.

\section{Ancestry and early years}

\textbf{Vovk:}
You were born in the North of England, in Blackburn, Lancashire,
but soon after your birth your family moved to London.
We know that some of your ancestors lived in Eastern Europe.
How did your family end up in Lancashire?

\textbf{Dawid:}
My maternal grandfather said he came from near Kyiv.
At the turn of the 19th and 20th centuries, 
there were pogroms against Jews in that area, and a lot left.
He was one of those who left.
He got on a boat, and he believed he was going to New York.
And when they got to Liverpool, they said this is New York.
And he got off [laughs].

I am not quite sure how he made the journey from Liverpool to Blackburn.
But he did end up in Lancashire and Blackburn.

My maternal grandmother had come from a little town in Poland.
I don't know a lot of details, but she certainly had quite a number of siblings.
They tended to settle around Manchester and Leeds, so I had a lot of relations in that area.
My mother was born in Leeds.

As for my father,
he was also escaping persecution, before the Second World War.
He was originally from a place in Ukraine,
but he told me it was Polish.
It was in Bukovina,
one of those areas where by standing still for a long time,
you could have five or six different nationalities.

His family moved to Vienna around 1920,
and he spent his teenage and formative years there.
He studied medicine there but left in 1938,
when there was a lot of Nazi persecution.
Various other members of his family,
including his mother and sister,
did not manage to get out and were taken to concentration camps.

About 25 years ago, after he died,
I was at a conference in Vienna,
and I tried to trace some details of my father there.
I went along to the Rathaus, the town hall in Vienna.
They were very helpful
and found a record card for my father from around 1935.
It gave details of his parentage and his address,
which I later went to look at.
But most interestingly, it said his nationality was Romanian,
which I would have never guessed [laughs].
At some point or other, the Romanians were in control
and probably gave everybody Romanian passports.
The name of the town was Otynia.%
\footnote
  {At this time Otynia is in the Ivano-Frankivsk region of western Ukraine.}
It was probably a little Jewish town.

\textbf{Shafer:}
Abraham Wald also was in Vienna and also escaped in 1938.
His family was from Transylvania and had Romanian nationality;
he was the only one of his family to escape.

\textbf{Dawid:}
Nice, our links with Wald\dots.
My father came from there to England in about 1938, just before the war;
so he managed to get out in time.
His brother had gone before him.
They were both interned in a camp on the Isle of Man as suspect aliens.
And eventually after the war,
as he was a doctor,
he was assigned to and got a job in a hospital in Blackburn.
There weren't many Jewish families in Blackburn at the time,
so he met up with my grandparents' family and ended up marrying my mother.
And I was born there in 1946.
When I was two or three, the whole family moved down to London.
So my formative years were spent in London.

\textbf{Vovk:}
And do you know why they moved to London?
Or any details of how it happened?

\textbf{Dawid:}
I have no information at all about that.
They never said why they left Blackburn and came to London.
They ended up buying a large, rather ramshackle, house in West Hampstead and lived on the ground floor.
There were two more floors,
which they let off to various tenants.
The house that we lived in was eventually,
after we all left many years later,
demolished, and a new multi-occupancy property was built on the site.
What I rather like about that
is they preserved some of the architectural features of the original house.
Although it was a new building, it was in some sense recognizable
as the house where I had used to live.

\textbf{Vovk:}
At some point you went to the City of London School.
How did you find it?

\textbf{Dawid:}
Yes, I went to the City of London School for Boys.
Before that, I'd been at the local primary school,
where, in fact, my aunt was a teacher and later became headmistress.
So that was all very much in the family.
And then from that I had to go for interview
and eventually got admitted to the City of London School for Boys,
which was down on the river, near Blackfriars.
It's still down on the river,
but it's moved to a new building a bit further away now.

What was my experience?
When you move from one place to another, things are different.
It seems I was doing very well
and was one of the clever clogs at my primary school,
and then you go to a selective secondary school.
And of course everybody is a clever clogs,
and you are no longer assured of a position near the top of the class,
and for a long time I was closer to the bottom of the class.
Including mathematics.
I remember the education was excellent in every way at the school.
There were three or four maths sets.
And I was always in maths set number two
because I wasn't good enough to be in maths set number one.

\begin{figure*}
\begin{center}
  \includegraphics[width=12cm,trim={9cm 6cm 1cm 5mm},clip]{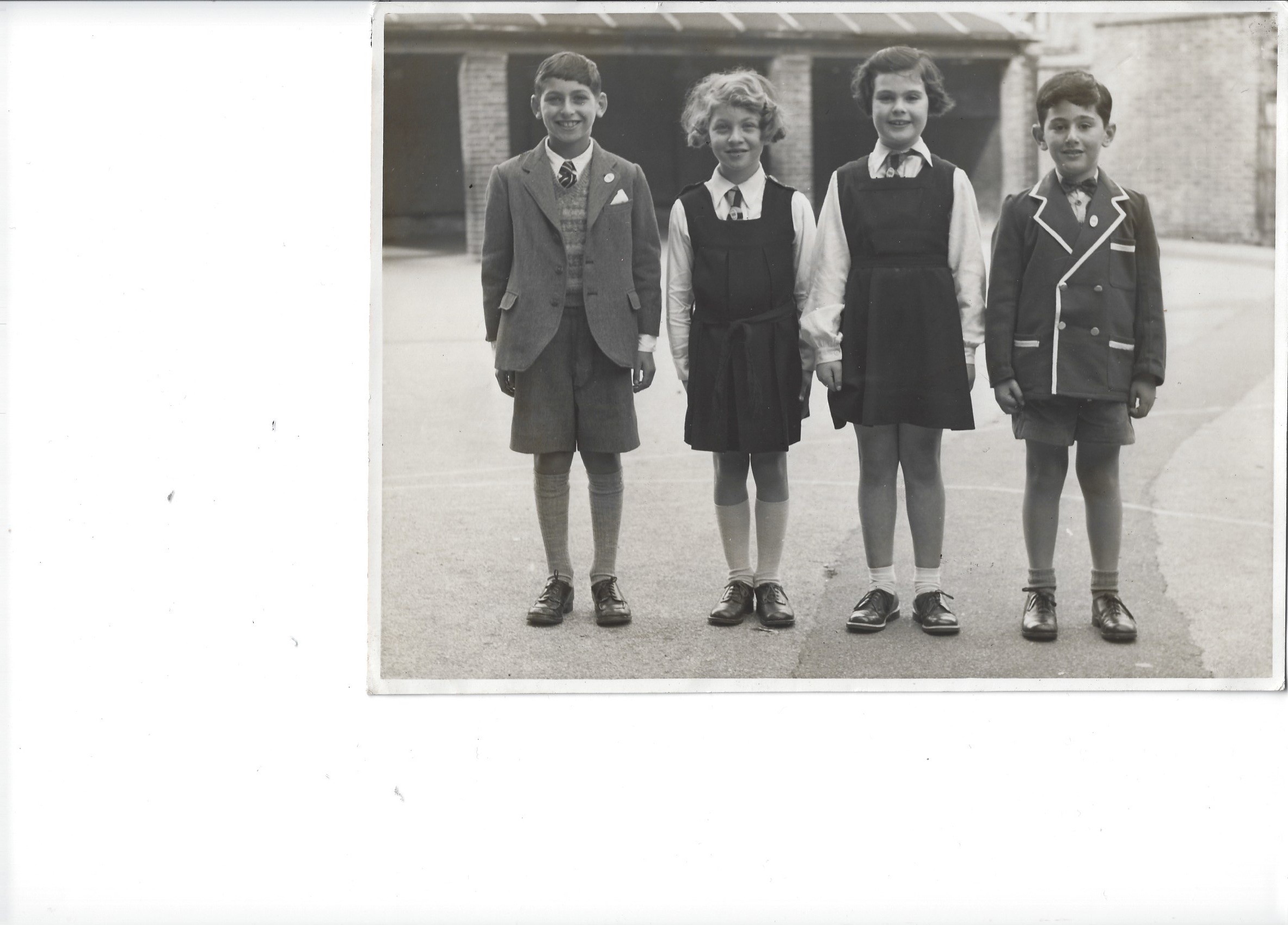}
\end{center} 
\caption{The winning road safety quiz team (ca 1955).
  Philip, aged ca 8, is on the right of the photo}
\end{figure*}

Not until the fifth form, coming up to O-levels,
did I really get to grips with maths.
And then suddenly it seemed to click into place.
For the first time since I don't know when,
somebody in maths set two came top of the whole year;
I beat everyone else.

The things I was enjoying at school most were actually the languages.
Ancient Greek, French, there was some Russian at one point.
I might have gone into the language stream,
but I ended up deciding to do mathematics,
going to the mathematics sixth form.
The teaching was splendid,
and at the time I knew nothing else.
I just thought this was normal and took it for granted.
And it wasn't till later on,
when I mixed with people who'd been at other schools,
where it wasn't at all splendid,
that I just realized how lucky I had been.
So I did nothing but mathematics and physics at A-levels,
very much concentrating at that point on the mathematical sciences.

\textbf{Vovk:}
Yes, your CV says you did A-levels
in Mathematics, Advanced Mathematics, and Physics,
so it was perfect for a mathematical subject,
and for statistics it must also be perfect.

\textbf{Dawid:}
Well, those were the days when we were very highly specialized.
Whatever you studied in the sixth form,
it was going to be very, very highly specialized.
And so in the mathematical sixth form, that's what we all did.

\textbf{Vovk:}
Did you miss an opportunity to do languages,
or biology, as your other A-level subjects?

\textbf{Dawid:}
Biology was never a big thing for me,
I do have to confess.
It was a strange thing:
as I said, we specialized so much,
and yet in some ways we didn't specialize.
So in the fifth form, when we were doing the O-levels,
we didn't have the option to do separate science subjects,
so we couldn't do physics.
I mean, these were available, but our school didn't do them.
So we didn't do physics, we didn't do chemistry.
There was a single science course called ``Biology and General Science''.
And it was incredibly superficial about everything,
so I didn't learn a lot from it.
I only remember, in the final examination,
we all sat down and, as well as the examination paper on our desk,
there was a little chunk of something white,
about half the size of a sugar cube and rather irregular.
And one of the questions was to identify this and describe it.
I didn't know if it was a tooth or a vertebra.
I don't know what it was [laughs].

\textbf{Shafer:}
I don't understand the English system so well.
So once you do this very specialized sixth form,
it's sort of taken for granted what you would study at the University.

\textbf{Dawid:}
Well, that left you very few options,
it has to be said.
At that point I couldn't go and study foreign languages at university.
Although in those days
(a lot has changed since then, the system has been overhauled),
I remember, when I had to apply for Cambridge entry,
I had to do entrance examinations.
You had a week of examinations, and that was quite broad.
Everybody who applied for Cambridge had to pass Latin, for example.
I did, remarkably.
I think it does not happen any more.

\textbf{Shafer:}
Could you summarize in what ways, if any,
your family, childhood, and early education shaped your intellectual interests?

\textbf{Dawid:}
My mother was a French teacher.
My aunt was a teacher and then headmistress of a primary school, as I said.
My uncle was a teacher.
My mother had three siblings, and three out of the four were teachers.
That was the family background;
nobody was particularly into the mathematical side of things.
They did help me into a public school,\footnote%
  {In modern British usage,
  public schools are fee-paying selective secondary schools.}
the City of London School for Boys.
I went there when I was ten as a scholarship boy,
which meant fees weren't payable.

\section{Higher education and teaching at Cambridge (then and now)}

\textbf{Vovk:}
You did BA in mathematics at Cambridge (Trinity Hall).
Did you enjoy it?

\textbf{Dawid:}
In my first year, we had a very good lecturer, Derek Taunt, in algebra.
The idea of abstract algebra was quite new to me.
I really liked
the understanding that you can manipulate all those patterns.
Although I've never been really tops in algebra,
I've got a good feeling for it.
I just like what you can do with symbolic manipulations.
What I didn't like very much was analysis.
I'm not very keen on things to do with limits and functional analysis.
Forget it.
And all these quantifiers I can't keep track of.
I generally was drawn towards the purer algebraic side of things.
We also had to do quite a lot of applied maths,
so there were courses in quantum theory,
for example, which I couldn't make head or tail of.
I've kept some of my notebooks from those days.
But you know, when I look at them,
unless it's something I've been doing since,
it really makes pretty no sense to me whatsoever.

\textbf{Vovk:}
I thought geometry or differential geometry might have impressed you,
because it's something I've always thought you were good at.

\textbf{Dawid:}
Well, geometry and algebra, of course, are very close in a way.
I mean they are distinct, but I regard them as close now.
In fact, going back to my secondary school,
we had splendid geometry lectures and lessons in my school.
So I love that a lot.
It was all very much ancient Greek stuff,
properties of triangles and things,
but it was a lot of fun.
We didn't have much in the maths syllabus at Cambridge
which is specifically geometry as I recall.
I'm trying to think if there was any.
I can't remember there being any course which was pure geometry.
As for differential geometry, that's a later chapter in my life.

\textbf{Vovk:}
Did you have any courses in probability or statistics in your BA programme?

\textbf{Dawid:}
There was a little, but it was little.
My first introduction was a course in the first year taught by John Kingman.
I had never heard of John Kingman,
and he was quite junior in those days anyway.
But he was clearly a very clever guy.
The lectures were all supposed to be 50 minutes long,
but he always finished about 15 minutes early
and then somehow had to extemporize
to try and fill the rest of the time [laughs].
But it was very basic indeed.
It was just axioms of probability theory
and maybe one or two little distributions.

We had a course in the second year called ``Random Variables'',
which was more concentrated on Poisson, hypergeometric, and the like,
introducing families of distributions,
with the occasional limit theorem as well.
I remember we had a couple of Borel--Cantelli lemmas.
Don't ask me now what it is.
I know it's a little theorem about zeros and ones
[laughs].

\textbf{Vovk:}
But no statistics?

\textbf{Dawid:}
Not much.
I don't think we got the t-test or anything in the third year.
I mean we had a lot of options in the third year,
and I can't remember if there were any statistics options;
but if there were, I didn't take them.
The statistics and probability content of the Cambridge Undergraduate course
has certainly grown over the years,
but it was incredibly minimal at that time.
But I do remember that there was one thing which ignited my interest,
which was hypothesis testing.
The very basic introduction to the Neyman--Pearson lemma.
I thought that's great.
There's actually a principled way to think about this problem and put it into mathematics.
I'm not sure I would have the same thoughts about it now,
but that actually attracted me, and maybe that's why later on,
when I had to choose what I was going to do after my first degree,
one of the things that came to mind as a possibility, not the only possibility,
was to do a course in statistics.

\textbf{Vovk:}
After your BA degree programme,
you did a nine-month Diploma in Mathematical Statistics.
It's a famous programme.
[See, e.g., \citet{Whittle:2002}.]

\textbf{Dawid:}
It doesn't exist any more, however.

\textbf{Vovk:}
It's a pity.
In a previous interview [\citet{Oliver:2019}]
you mentioned that you became a Bayesian after attending lectures
by David McLaren.\footnote%
  {David McLaren joined the Statistical Laboratory from Manchester as research student.
  David Kendall, the first Professor of Mathematical Statistics in Cambridge
  and \emph{ex officio} Director of the Statistical Laboratory,
  secured his appointment there
  as a senior Assistant in research
  (later he became an Assistant Director of Research),
  with partial responsibility for statistical consulting
  \citep{Whittle:2002}.
  After Cambridge he moved to the University of Glasgow.
  He was the seconder of the RSS discussion papers
  by \citet{Dawid/etal:1973} and \citet{Dawid:1979}.}

\textbf{Dawid:}
I remember in the first week or two on my diploma course,
when we were first introduced to this idea of the Bayesian approach,
I thought ``what absolute rubbish'';
how would anybody possibly think about things this way?
But then David McLaren gave his course, and it was very interesting.
He was quite junior,
but he was one of the most thoughtful guys I've ever met.
And he gave the most fascinating course,
which was called ``Practical Statistics''.
This was a course that had been given for decades at Cambridge.
It was about learning to do chi squares and t-tests
and computing them.
You go over little numerical problems,
and you have to essentially crank out the t-test or something.

This was given to him as one of the junior members of staff to teach.
But what he did was he took the various problems that had been handed down,
and he started thinking about them.
He would get up in front of the class,
would say here's this problem,
things like the bus number problem.
You find yourself in a strange city,
and the first three buses that pass you are numbers 3, 7, and 21.
And the question is what can you say about how many buses there are in the city?
So this is sort of question you were asked,
and then you had to think about modelling it.
He took it very seriously;
how do you model these problems
based on the opinions you had had before you saw them?
He introduced the Bayesian approach
entirely as a practical methodological approach
for tackling problems.

He didn't give us any philosophy of it,
but it really kindled an interest in me.
I thought, yeah, this is fun,
this is definitely a good way of going about things.
And so that was very formative.
And then I also remember,
besides the lecture course he was giving,
that at one point he gave a seminar,
and in the seminar he presented Birnbaum's proof of the relation
between sufficiency, conditionality, and the likelihood principle,
and I thought, wow, this is amazing.
And so little by little I was really hooked
on the more philosophical and logical issues underlying statistical inference,
which has been a major theme ever since.

\textbf{Vovk:}
How would you describe the kind of Bayesian statistics
McLaren was doing in his course:
was it mainly about using the Bayes theorem,
or did he also discuss details of building Bayesian models,
such as personal probabilities elicitation?

\textbf{Dawid:}
Yes, very much so.
He would take one of these problems that had been handed down over the decades,
and he was saying let's think about it.
Let's think about what are we trying to do,
and what do we know, and how are we going to describe it.
And so first of all we start thinking about the modelling aspects,
what do we know, what don't we know,
and what do we think we know about what we don't know.
So we started thinking about prior opinions.
It was the whole construction of a Bayesian analysis for a ``real'' problem.
Not truly a real problem, of course; they were toy problems.
But problems none the less.

\textbf{Vovk:}
So it was much wider than just using the Bayes theorem\dots.

\textbf{Dawid:}
You needed to know how to do it, but that was the most trivial part of it.
The Bayesian machinery is just handle turning.
But you need to know what to feed into the machine.

\textbf{Vovk:}
And how did your attitude about the Neyman--Pearson lemma change?

\textbf{Dawid:}
Of course, that came back several times in various forms.
I understood there was something interesting there,
but what I really didn't like was this business of the five percent or $\alpha$,
setting it,
fixing one kind of error and letting the other look after itself,
all that seemed a bit of a strange thing to do.
And of course, it's not the only thing you can do;
there are Bayesian alternatives.

When I was teaching my Cambridge course to the undergraduates,
I thought of the Neyman--Pearson lemma
as one of the prime examples of what I call
the fundamental theorem of Bayesian inference,
which is the equivalence of normal and extensive approaches to analysing problems:
whether you think of an optimal strategy before you've seen the data
or an optimal act after you've seen the data.
And of course if you're a proper Bayesian and do things right,
you get the same answer.
If you do that, you can see the Neyman--Pearson lemma
falls out, with the likelihood ratio test as the only thing that makes sense,
and you can talk about decision theory and complete class theorems,
and you've got to do a likelihood ratio test.
The only question is which one,
and the Neyman--Pearson approach, which said we choose it by fixing $\alpha$,
I thought was a pretty stupid way of deciding which one.
And the Bayesian approach,
which balances prior probabilities and losses
and essentially says to find a cut off
and look whether your likelihood ratio is above or below the cut off,
where the cut off is externally determined,
that's the obviously sensible thing to do.
So the Neyman--Pearson lemma
is actually an excellent example of the fundamental theorem of Bayesian inference.\footnote%
  {See, e.g., \citet[Section 6.1, especially Proposition 6.1,
  and Section B.3.3]{Bernardo/Smith:2000}.}

\textbf{Shafer:}
Can you repeat some of that, Philip, what you said?
Fundamental theorem of Bayesian inference, is that what I heard?
And could you state that for me?

\textbf{Dawid:}
Let's talk about Bayesian decision theory.
Here you are going to do an experiment.
You're going to observe some data.
You've got some decision structure,
some loss function depending on your terminal action and the state of nature.
Now there are two ways to approach that.
One is before you do the experiment;
you can consider various decision rules and try and choose among them.
Each of those decision rules will have an expected risk, Bayes risk.
And so the normal form of analysis tells you
to choose the decision rule with the smallest Bayes risk.
And then after you've done the experiment,
you find out what the outcome was,
and you plug it into the rule you've chosen,
and that gives you your preferred action.
The extensive form of analysis says:
well, I've done my experiment, I've got my data,
I've updated all my probabilities because I'm a Bayesian;
and now I could choose between acts
according to which made the smallest expected loss.
So instead of choosing a decision rule before I see the data,
I choose an optimal act after I've seen the data.

And the fundamental theorem of Bayesian decision theory says
they give the same answer.
So the Neyman--Pearson lemma is basically looking
at the normal analysis of the two-hypotheses problem.
You're choosing a decision rule.

The way to solve it is to do the extensive analysis,
and you see immediately that it has to be based on the posterior odds,
which is essentially the likelihood ratio because the prior odds are fixed.
So that's the best way to prove the Neyman--Pearson lemma.

And of course it matters how you prove the Neyman--Pearson lemma,
because if you start worrying about 5\% or $\alpha$ levels,
then you generalize to more complicated problems,
composite hypotheses, and everything.
Then you're anchored to that way of thinking about it.
Whereas if you start from the Bayesian understanding of it,
you have a completely different way of generalizing it.

\textbf{Vovk:}
When did you realize it, Philip?
While you were still a student, or was it later?

\textbf{Dawid:}
It's an interesting question.
There was a lovely book that Dennis Lindley wrote.
[See \citet[pp.~13--14]{Lindley:1971}.]
So it was when I already was an academic with Lindley,
in my early days there.
He developed the argument together with Savage.
It is a very simple argument for why, in the framework of the Neyman--Pearson lemma,
it is rational to minimise a given linear combination of $\alpha$ and $\beta$,
but not (for example) to prespecify~$\alpha$.
At an early conference in my career,
I think a Royal Statistical Society Conference,
I actually gave a talk (there was no written version of it),
which I called the ``Voyage around the Neyman--Pearson lemma''
and which discussed all these kinds of things.

\textbf{Vovk:}
It's a pity it wasn't published.

\textbf{Dawid:}
Let me try to remember what I said.
Very vague memories.
I may well have mentioned Lindley's argument above,
and also the argument from ``the fundamental theorem of Bayesian decision theory''.
I think I also introduced a weaker form of the ``law of likelihood'':
evidence $E$ supports hypothesis $H$ over hypothesis $K$ more than evidence $F$ does if
\[
  p(E \mid H) / p(E \mid K) > p(F \mid H) / p(F \mid K).
\]

\textbf{Vovk:}
Do you remember any other people at Cambridge
while you were doing your Diploma?

\textbf{Dawid:}
Oh yes, we had very good teachers.
The head of the statistical laboratory at the time was David Kendall,
and he was amazing.
I remember he gave a course on Markov chains,
and it was incredibly advanced and abstract;
it was all disintegrations and Hilbert spaces and stuff like this.
He had this amazing way;
you sat in the lecture and soaked it up,
and you understood absolutely everything he said.
And the moment he finished, you had no idea what he was talking about [laughs].
Because it was just way above my head, really.
But he had this wonderful way of making you see it as simple.
For him it obviously was.
For the large part,
he taught very pure probability theory,
but he also was interested in other things.
There was an opportunity for the teachers there to give courses
on their own special interests,
and he had just got interested in modelling bird flight.
And so we had a course on modelling bird flight.

There were plenty of other interesting people.
Maurice Walker gave us the main deep theoretical statistics course.
And the content was wonderful.
Even though he had an incredibly dry way of delivering it.
He did a very technical analysis of, among many other things,
asymptotic normality of posterior distributions,
with all the detailed technical conditions and everything.
And my first published paper [\citet{Dawid:1970}] was an extension of that;
it was built on the theory that he'd given us there.

There was Bob Loynes,
who had taught me as an undergraduate the course on random variables.
The course he taught in the Diploma programme
may have been on Applied Probability.
It was clear that he'd been assigned a course he wasn't really tops in.
And that's happened to all of us.
And he was maybe staying one lesson ahead of us.

We got experimental design from Bob Bechhofer,
an American who was visiting us;
I remember that was good.
Overall, the teaching was excellent.

\textbf{Vovk:}
You had both probability and statistics.
What was the proportion between those approximately?

\textbf{Dawid:}
I think it was probably more loaded on the statistics side,
but there was a substantial amount of probability.

\textbf{Vovk:}
I know Cambridge calls it ``Statistical Laboratory'',
but people are doing mostly probability there.
It may have changed recently.

\textbf{Dawid:}
There's a mixture there.
There are subgroups within the lab.
There is a section which does theoretical statistics
and a section which does pure probability
and a section which does operational research.
The strengths probably have wandered a little randomly
over the course of time,
depending on personnel and interests.

\textbf{Shafer:}
Can you imagine putting yourself
in the shoes of an undergraduate in mathematics at Cambridge today?
Do you think you would go into statistics?

\textbf{Dawid:}
When I was in Cambridge as professor,
I was given a second year undergraduate course,
and I thought it was great,
because it was called ``Principles of Statistics''.
And as you know, I'm very keen on principled thinking.
So I looked at the syllabus for this and the notes from previous lecturers,
which were almost chiselled in stone;
you weren't really supposed to depart very much
from a very clearly organized and specific syllabus.
And I couldn't see one principle in it from beginning to end.
It was the usual t-tests and chi squares.
And a bit of some theory including Neyman--Pearson.
I rebelled a little and insisted on inserting a small section
on conditionality and likelihood and stuff like that.
I think some of the students quite liked it.
But I got disapproving looks and comments from my colleagues
because that wasn't really the sort of stuff that was supposed to be there.

\textbf{Vovk:}
Not mathematical enough?

\textbf{Dawid:}
Well, it just wasn't the standard material,
and it was a bit too philosophical, perhaps.
It was mathematical in its way, but it wasn't just mathematics.
It was mathematics that you had to think about.
It was the right mathematics.

\textbf{Shafer:}
But as a student, if you were in that environment,
there are so many other options
as compared with your time\dots.

\textbf{Dawid:}
But I can tell you what I think I would like to have done,
and it's not a mathematical subject.
I would like to have done something on genetics.
There were some bits of probabilistic population genetics
that we had in the course, but very little.
Because modern genetics is such an enormous field
with so many directions.
I thought I had a grasp of and a gift for doing maths, and I knew I could do it well.
I know that if I started genetics,
I wouldn't have done well, and I'd have sympathized with those students
who were at the bottom half of the class and struggling
but really, really fascinated.
Because I find it challenging, very challenging.

Going back to the mathematics programme,
where would I have gone as a mathematics undergraduate now?
A lot of them did carry on and do statistics.
I said the Diploma was sort of retired,
but it lived on in the form of what they call Part III.
There was a change of name and some change of substance,
but largely it lives on,
and you can do courses pretty similar to what you always did on the Diploma.

But as you're saying, there're so many other strands.
There were two postgraduate diplomas in my time:
as well as statistics, there was the Diploma in Automatic Computing.
What would then pass for computer science.
But that didn't attract me.
It was kind of interesting, but it didn't attract me quite so much.
That was all that was available at the postgraduate level.

There is such a strong emphasis in Cambridge,
because it is now part of pure mathematics department,
on the pure side of maths,
which in some ways I like,
that maybe I'd have ended up doing something there.
I think one of the reasons I went into statistics,
without knowing much about it,
was because I thought maybe there would be job opportunities.
Pure maths was a lot of fun, but how do you capitalize on it?
How do you monetize it?
I never did discover how to monetize statistics,
because I ended up in academia anyway,
but that was the original motivation [laughs].

\textbf{Vovk:}
It's a recurrent theme in \emph{Statistical Science} conversations\dots.
I remember this is what happened to Doob.
I think lots of people felt statistics was more practical
and more promising.

\textbf{Dawid:}
You need to go out in the big wide world and make a living.

\textbf{Vovk:}
There is a beautiful set of notes called ``Realized path'' by Peter Whittle
about the Statistical Laboratory in Cambridge
[\citet{Whittle:2002}].
In it he writes that the Cambridge Diploma in Mathematical Statistics
had remained true to its initial conception
that the theoretical grounding should be accompanied by close acquaintance with an applied field
and a testing investigation of data from that field.
What was your field if you had one?

\textbf{Dawid:}
Yes.
In the Diploma, when I took it, the idea was that we had to do an applied project,
often with somebody from a different department.
So I was teamed up with somebody in a department with a strange name;
it was called the Department of Human Ecology.
There was a statistician there called Bob Carpenter,
who later went to the London School of Hygiene and Tropical Medicine.\footnote{%
  Bob Carpenter was in the Department of Human Ecology at Cambridge University from 1961,
  and in 1971 he became Senior Lecturer in Medical Statistics
  at the London School of Hygiene and Tropical Medicine.}
He was in Cambridge at the time, and I started doing a project with him.
I can't quite remember what it was originally meant to be.
What it rapidly turned into was the following.
At that time they were just completing the building of a new hospital.
Addenbrookes Hospital had been on a central city site,
and a new hospital was being constructed on the outskirts of the city.
It is now an enormous medical campus.
And there was one small problem, which was the following.
They wanted to install laundry chutes so that the nurses on any floor would open the chute and put in the bag of laundry,
and it would drop down.
And the question is how should they design it to minimize the chance of two bags jamming the chute.
And this was my project [laughs].
It's fascinating really.
I got details of the design,
and I wish I could actually find what I wrote.
It wasn't really long.
I don't know if anybody took any notice of it,
but I had to think about it, and we had to do some experiments.
The question was: in the chute, can we regard these things as falling under gravity as if unconfined?
So we actually did some experiment with dropping things down and timing them,
and the answer was essentially yes, they just dropped\dots.
That's what I remember.
Don't ask me what my final recommendations were because I can't quite remember,
and please don't ask me what that had to do with statistics.

Since the diploma was a nine month course,
you had to fit the project in during all the lectures.
I would go every week or two, I think, to talk to Bob Carpenter
and to go along to the hospital and see how things work.

\textbf{Vovk:}
After you earned your BA degree in Mathematics and Diploma in Mathematical Statistics,
you were supervised by David Cox at Imperial for a year.

\textbf{Dawid:}
That's right.
I left Cambridge,
and I didn't quite know what I was going to do,
whether I wanted to do a PhD or not.
Now, I did very well in the Diploma exam.
And then David Kendall suggested that I might stay on,
but I didn't respond to that immediately,
and I thought about it for a while.
Then I got in touch with him and said, yes, I'd be interested.
And he said, ``I'm sorry, the funding has gone to somebody else'' [laughs].
So that was out.

But having thought about doing a PhD,
I thought, well, where could I do it?
And I had a relative who had been to courses of Cox.
Through that I made contact with Cox,
and I became his PhD student at Imperial College.
But we had a big mismatch in our interests.
I really wanted to do something Bayesian,
and he really wanted me to do something with things he was doing
of the Neyman--Pearson type for testing separate families of hypotheses.
We just weren't seeing eye to eye.
I did at that time, essentially self-propelled,
write my first publication,
which I already mentioned,
on asymptotic normality of posterior distributions.

I was looking around,
as it clearly wasn't the environment I wanted to be in.
And I'd come across Lindley's wonderful two-volume
``Introduction to Probability and Statistics from a Bayesian Viewpoint''.
[See \citet{Lindley:1965}.]
It's a bit out of date,
but still wonderful, beautifully written, and that had impressed me enormously.
  And then I realized Lindley had just been appointed
to the Chair at University College London.
I made contact with him,
and my publication either had come out, or it was available,
and he'd seen it.
In fact, I have an idea he was asked to referee it.
And so he decided he would take me on.

\section{UCL, City, and back to UCL}

\textbf{Dawid:}
So I moved just a short distance across London from Imperial College
to University College,
and I spent a year with Lindley.
We got on very well, except we didn't see a lot of each other.
I got on very well with everybody at UCL.
It was a very conducive environment to being a Bayesian,
especially with Mervyn Stone there,
and he was a tremendous influence on me.
As for Dennis Lindley,
we sort of fooled around trying a few ideas I might make a thesis out of,
and nothing ever really came of it.
Most of the time I just kept out of his way.

Towards the end of that year
a conflict arose with the previous regime,
which was under Maurice Bartlett.
I think there was a bit of personal as well as academic friction
between Lindley and some of the old timers,
because they weren't seeing eye to eye in many ways.
Quite a few of them were leaving.
So one guy left,
and basically Lindley just gave me his job.
He said, ``Phil, would you like this job?''
In those days,
there was no advertisement, no appointment process, no interview,
no oversight, no nothing.
He just gave me the job, which was nice [laughs].

So keeping out of his way was obviously the right thing to do.
I did one more year officially in the PhD,
although I didn't really make much progress with it,
and then I started as a Lecturer.

\textbf{Shafer:}
You would use this term,
``keep out of his way''.
Was it accidental, or was there some reason?

\textbf{Dawid:}
No, we wouldn't avoid each other.
No, not at all.
But I didn't have many supervision sessions and sort of went off on my own.
It was a very formative year for me.
I didn't actually produce much and did an awful lot of reading.
I particularly remember being impressed with Ferguson's book
on decision theory [\citet{Ferguson:1967}].
It was a wonderful book.
Still is, and I remember soaking that up and things like that,
so I sort of was stocking up.
Putting fuel in there that I would burn later.
Getting the nutrition that I would then build on [smiles].

\textbf{Vovk:}
And how would you compare it with your year at Imperial?
Did David Cox give you any problems to solve?
What was his style of supervision?

\textbf{Dawid:}
Cox didn't give me any problems I wanted to solve.
At one point, I seem to remember,
there was a problem on the behaviour of woodlice.
Apparently, if you put woodlice on a certain surface,
they go around various interesting tracks,
and there was a connection with where the moisture was and things like that.
And I actually started quite seriously thinking about it
before I realized I didn't really care about woodlice [laughs].
So I gave that up.

No, Cox didn't give me any problems I wanted to work on,
and I came along with problems I wanted to work on,
like posterior normality,
and he wasn't the least bit interested in that,
so it wasn't a fruitful relationship.

It was a bit of an isolated existence.
You didn't really meet people very easily in that environment.
There was a large building on Princes Gate in South Kensington,
opposite to the main site of Imperial College,
a building which was converted for the use of the statistics department,
and there was a room in that which was basically a boardroom
with a very large table and with lots of chairs around it,
and a piano in the corner which people could book to practise piano.
And three research students were assigned to this room.
There was myself, John Fox, and Peter Bloomfield.
And we didn't really have any reason to be there
when we didn't have a supervision session with Cox.
So we went in one day a week.
And they were different days,
so it was like Box and Cox sharing a room.%
\footnote{According to the Oxford English Dictionary,
  ``Box and Cox'' (as noun in British English) means
  ``A situation or arrangement in which two or more people take turns
  in occupying the same space or position,
  sustaining a part in some activity, etc.''
  It derives from an 1847 farce by John M. Morton
  and its two characters, John Box and James Cox.
  George Box discusses the Box and Cox story in \citet[p.~254]{DeGroot:1987}
  as the source of his joint paper with David Cox.}
It wasn't till about three months that I actually bumped into John.
And then another day I bumped into Peter,
and finally I was able to introduce them to each other [laughs].

So it was a very strange existence.
Whereas back in University College,
it was much more like normal people interacting in fairly normal ways.
Everything had to stop for a four o'clock tea,
and the secretary would bring in cucumber sandwiches.
So we would socialize.
That was a very different environment.

\textbf{Vovk:}
And you were working on your own problems\dots.

\textbf{Dawid:}
Well, in the year I was officially doing a PhD, I wasn't.
As I said, I was doing more reading than producing anything.
Later I started lecturing, and I took that very seriously.
So I wasn't thinking about research for some time.
Eventually I went to see Dennis Lindley, and I said,
``Look, I need some pressure or impetus to finish my PhD.''
``Oh'', he said. ``Don't bother about that, Phil, just write papers.''
And of course,
Dennis Lindley never had a PhD himself.

\textbf{Vovk:}
And maybe it's true about Fisher as well;
in his books he used post-nominals ``ScD, FRS''.

\textbf{Dawid:}
There were reasons of war years and things like that for a lot of people.

Pretty soon after that I got really interested in stuff that Mervyn Stone was doing.
An example of our early joint work is
``Expectation consistency of inverse probability distributions''
[\citet{Stone/Dawid:1972}].
That was a look at the logic of whether inverse probability distributions
after seeing the data should look like Bayesian posterior distributions,
and what principles you could put there.

My first joint paper with Mervyn (and second overall) was
``Un-Bayesian implications of improper Bayes inference in routine statistical problems''
[\citet{Dawid/Stone:1972}],
and that was the first time we talked about the marginalization paradox.
Actually I remember the marginalization paradox arose from my teaching.
I was teaching a Master's course on statistics,
where I was, unlike Cambridge, free to do almost anything I wanted.
So it was quite a Bayesian thing.
And then I had to set the examination, and I set one question,
which was something where I,
just by fooling around with a particular problem,
actually discovered the marginalization paradox.
And I set it as an exam question.
So it first came to light as an exam question
from the Master's course,
which the one student on the course failed to answer [laughs].

And then I started discussing with Mervyn,
and we wrote a little paper on it, which had the basics of it,
but not the underlying theory.
That was developed when the next year Jim Zidek
came visiting from Vancouver, BC,
and he had the mathematical skills and understanding to put it in place.
And that became the Dawid--Stone--Zidek paper ``Marginalization paradoxes''
[\citet{Dawid/etal:1973}],
which we carried on thinking about for many decades later.

\textbf{Shafer:}
Did you feel the heritage of University College,
being the first stat department?

\textbf{Dawid:}
In a way, yes.
There was a certain pride, shall we say,
knowing that this was where it basically got started.
Although of course, there was very little continuity
between the sort of stuff that was done in the early days
and what Dennis Lindley and his crowd were doing.
But we did feel proud to be in this ancient founding university for statistics.

\textbf{Shafer:}
Was the afternoon tea a tradition going back to Karl Pearson?

\textbf{Dawid:}
It probably was actually, yes.

\textbf{Shafer:}
So tell us about Mervyn, how did he get there?
Did he come with Lindley?

\textbf{Dawid:}
Yes, he came a year later.

Lindley was appointed to the Chair at Aberystwyth.
He'd been in Cambridge, and he'd been Director of the Statistical Laboratory,
but they never appointed him there as the top professor.
And so to get a chair, he moved to Aberystwyth,
which was very much out of the way.
And Mervyn Stone also ended up in Aberystwyth.
I think he'd been at Durham originally.\footnote%
  {The timeline \citep{Galbraith:2021} is:
  In 1961 Mervyn Stone was appointed to a lectureship
  in Lindley's new statistics department at Aberystwyth.
  After spending the year 1965--1966 at the University of Wisconsin,
  he took up a Readership at the University of Durham
  and in 1968 moved to UCL as a Reader (later Professor) of Probability and Statistics.}
So Mervyn and Dennis were together.

Dennis had Mervyn in his department in Aberystwyth,
and there were others.
Dennis came and Mervyn followed,
and also Rodney Brooks was another one
who followed Dennis from Aberystwyth.
So he had a rump of a department to build on.

Mervyn was such an original guy.
He's certainly the most original guy I've ever met.
He had a completely different way of thinking about things
from anybody else I know.

\textbf{Shafer:}
Can you elaborate on that?

\textbf{Dawid:}
It showed positively,
but also negatively as well in later years.
When he got to the stage where he became actually entirely incomprehensible.
There was this guy here with absolutely wonderful ideas
and quite unable to express them in a way that anybody else can understand.
That may have been because the ideas were simply ineffable.
So it wasn't possible for mere mortals to understand them.
And I think that was part of it actually.
But also it was very frustrating.

\textbf{Vovk:}
Was your move to City University
connected in any way with Lindley's early retirement?
He talks about his early retirement in 1977
in his \emph{Statistical Science} conversation [\citet{Smith:1995}].
It is close in time to your move to City in 1978.

\textbf{Dawid:}
I don't think I moved because Lindley had retired.
In fact, after Lindley retired, Mervyn Stone became Head of Department.
And you know, we were very close buddies.
I wasn't being pushed out of UCL in any sense,
but an opportunity had come up at City University.
In fact, the chair was vacant,
sadly, because Allan Birnbaum had been the previous occupant,
and he committed suicide in the department.
So there was a vacancy,
and I applied for that.
Before City, Birnbaum had come over as a visitor to UCL for a year from America,
and he was working on his likelihood stuff.
He was trying to write a book.
So we interacted, and then a vacancy came up at City,
and he got the chair of City University.
But he didn't last long,
and for various reasons, mostly personal, I think, but perhaps also academic,
he committed suicide.
This is very sad.

There was a vacancy, which then I filled.
And so I went for three years to City University,
which is again a very different environment.
It was not a top ranked institution,
although there were some very good people there.
But there were also some mediocre people.
I really didn't want to hang around there too long.
I could have left after two years,
but I thought they might have problems filling the post.
So I stayed on one more year,
and they had a real problem filling the post actually:
appointed somebody and then revoked the appointment
because of financial reasons.

But anyway, that's really beside the point.
So I got the opportunity of going back to University College,
but as a demotion.
I'd been a professor at City University,
and the only opening at UCL was for a Readership,
which, as you know, is one step down.
But I was essentially promised without being promised
that I would soon be promoted.
I couldn't be given an absolute definitive promise,
but it was made clear it was going to happen.
And it did happen.
So I was a Reader there for a year,
and then I was promoted back to Professor.

\textbf{Shafer:}
In his \emph{Statistical Science} conversation,
Dennis Lindley says he retired because he was made some financial offer
by the university.
But was it the only reason?
Do you think maybe he was tiring of his administrative duties?

\textbf{Dawid:}
I can't remember what he himself said,
but my memory of it,
which is second-hand, of course, maybe even third-hand,
is as follows.
The department he took over changed a lot when we took on a whole load of computer scientists.
We became a Department of Statistics and Computer Science.
And that wasn't a very happy marriage.
Lindley started looking for ways
he could legally get rid of some people [laughs].
And he started looking through all kinds of documents,
things like the university rules and regulations,
about how we could encourage their early retirement,
and in doing so he realized that he could take early retirement [laughs].
It was financially beneficial.
So that's what he did
and thus succeeded in his desire to get away from these people [laughs].
I think he got quite a good financial arrangement out of it.
He was very happy to resign;
he was in his early 50s, I think.

\textbf{Shafer:}
So you wouldn't characterize it as retirement out of frustration
that he couldn't hold the department the way he wanted to.

\textbf{Dawid:}
Not so much on the statistical side.
On the statistical side,
even though we weren't quite as gung ho Bayesian as Dennis would like us to be,
we were a good, nice, compact, generally good-natured community.

\textbf{Vovk:}
At City University you were Head of Statistics
and Director of Statistical Laboratory.
Did you have a heavy administrative load?
Did you have enough time for your research?

\textbf{Dawid:}
Yes, I did a decent amount of research.
There was a lot more admin stuff at City University
in the sense that they just loved having meetings.
There were all sorts of committees
and things most of which had no effect on anything [smiles].
But they gave people something to do
and reasons to argue with each other.
I hated these meetings.

\textbf{Vovk:}
In 1982 you were awarded an ScD (Doctor of Science) degree.
You held top positions at City University without it;
was it required at UCL?

\textbf{Dawid:}
It was not required,
but it would be nice to be able to call myself Doctor,
that's all.
So as I recall, I first put in for it
when I was in my last year at City University.
The system was that if you were a graduate of Cambridge University,
you had the right to put yourself in for this degree by publication,
and it was really a question of making a bundle of publications.
And sending them off and waiting for a year,
to see if they met the grade or not.
And apparently they did.
So I got awarded the ScD.
And the only difference it made to my life really
was, when I did go back to Cambridge in more recent years,
as an academic and Fellow I was able to wear
the scarlet robe of the ScD [laughs].

\begin{figure*}
\begin{center}
  \includegraphics[width=12cm, trim={15cm 11.5cm 1cm 2mm}, clip]{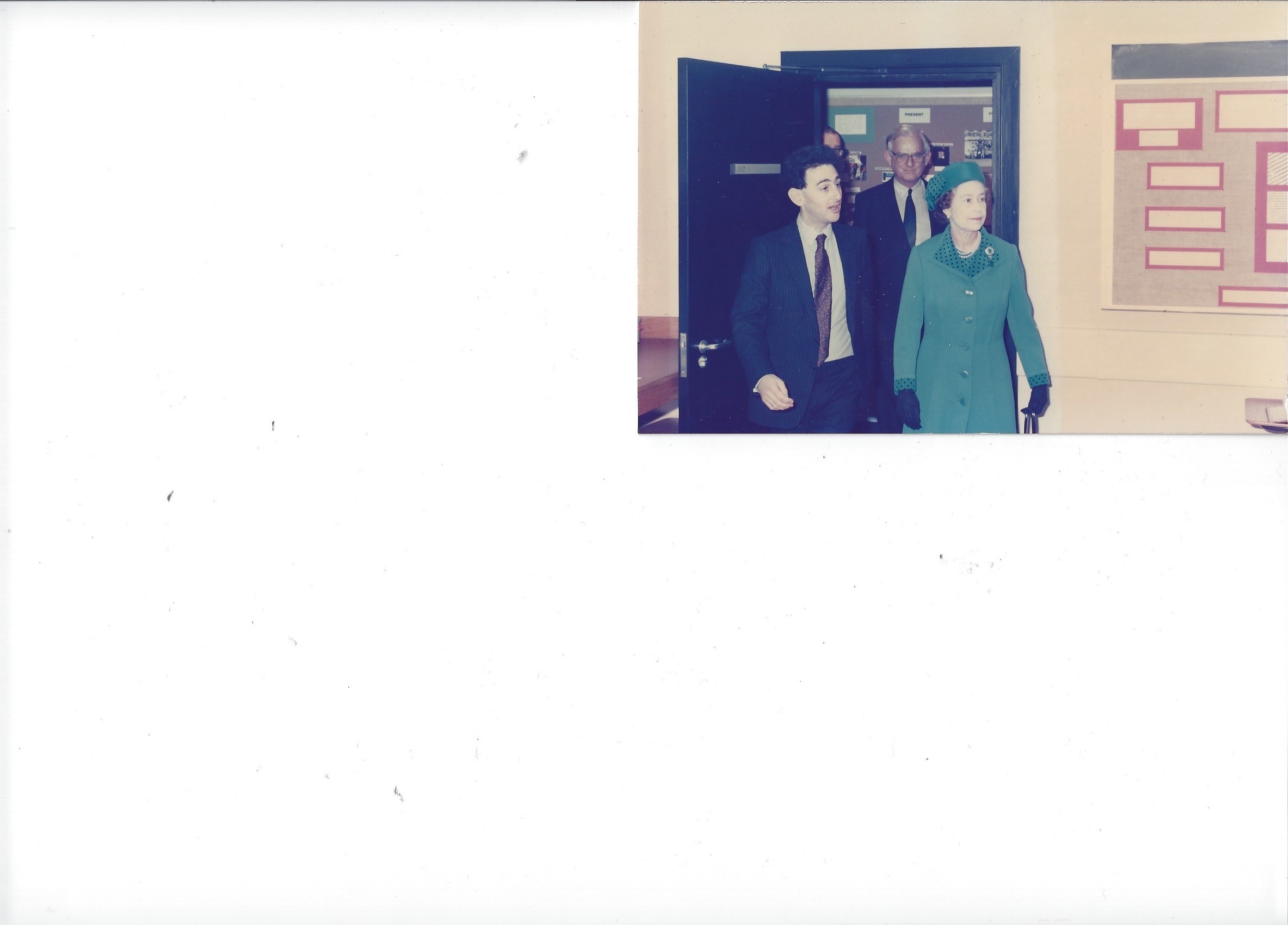}
\end{center} 
\caption{Showing Her Majesty Queen Elizabeth II round the UCL Statistical Science Department in 1985}
\end{figure*}

\section{Back to Cambridge}

\textbf{Vovk:}
Let's talk about your time in Cambridge as a professor.
We have touched on your time in Cambridge several times already,
but we mainly talked about your experiences as student.
In your interview with Alison Oliver you made some very interesting comparisons
between Cambridge and UCL.
At UCL statistics was important per se, of course;
it was the founding department of statistics.
And in Cambridge it was part of a bigger grouping.

\textbf{Dawid:}
Yes. In Cambridge it was a small corner of a very pure mathematics department.
And that did rub off on the kind of statistics that was regarded as worth doing.

\textbf{Vovk:}
Did this pull you in different directions?
Did you change the kind of research you were doing
when you moved to Cambridge?

\textbf{Dawid:}
No, I didn't change it and kept doing things I'd always been doing.
But there wasn't much empathy, much interaction
between the sort of things I was interested in and wanted to do,
and what most of the other members of the department wanted to do.
Apart from the postdocs and people I myself brought in,
Peter Whittle was the only permanent member I interacted with.
He'd retired, but he still often came in as an emeritus professor,
and we had some interesting discussions now and again.

\textbf{Vovk:}
Do you think either of these styles (UCL vs Cambridge)
is more conducive to the development of statistics,
or are both equally important?

\textbf{Dawid:}
I've got a very personal view on it,
which is that what is really important in statistics is logic rather than mathematics.
Philosophy even.
I've always been concerned about trying to understand the nature of things.
And what are we really trying to do here?
Mathematics is just a tool.
And most mathematical statisticians probably take a lot for granted
and then just do the wonderful mathematics
and important stuff with important theory and important applications and all the rest.
But very few share my own concerns with digging into what does it all mean and what should we be doing?
How should we be doing it?
How do we set the directions?
A lot of these directions have already been set.
Let's do some wonderful mathematics within those directions.
But that's me.
I find myself actually quite out of kilter with a lot of modern mainstream work in statistics
that's passing me by.
And that probably reflects on me rather than it does on the rest of statistics [laughs].

\textbf{Vovk:}
Do you remember any interesting work on the foundations of statistics done in Cambridge?
And did anybody share your interest in foundations?

\textbf{Dawid:}
I think it's fair to say that the only other person
who had any sympathy with the kind of thing I was doing was David Spiegelhalter,
who has been a very long standing friend and colleague.
He was appointed just about the same time as me
to this very interesting position of Professor of the Public Understanding of Risk.
He had lots of very applied things to do,
communicating to school children and stuff like that.
But he also did care about what he was talking about and getting it right,
and we did share some common understandings.
Even though we were doing things in very different ways.
In fact, we're still working together.
Only yesterday I was looking at something which is a joint project
to do with preparing an online course,
``Statistics for forensic science''.
You know, there are very interesting, subtle, logical and philosophical points there.
Forensic inference is fascinating, and it's got some interesting logical puzzles, subtleties,
and it's quite hard to put across to those people who need to know it.

\textbf{Vovk:}
Coming back to the difference between UCL and Cambridge styles,
do you think it showed in any way in the styles of teaching?
Were the students affected by it,
or were they taught essentially the same things in both places?

\textbf{Dawid:}
The Cambridge syllabus was definitely much more mathematically focused.
There's much more emphasis on rigorous deep mathematics than at UCL,
which also reflects a very different student body.
The students who went to Cambridge were essentially top mathematicians,
whereas the ones who went to the UCL Department of Statistical Science
probably liked maths,
but they weren't very good at it,
and they thought, wrongly, that statistics would be easier.

\textbf{Vovk:}
Did you have any interesting discussions with machine learning people at Cambridge?

\textbf{Dawid:}
When I was still at UCL,
Geoff Hinton,
who's recently won all sorts of wonderful prizes for deep learning and all that sort of stuff,
had recently come to head up the new Gatsby Institute for Computational Neuroscience,
which was machine learning.
And shortly after that, he arranged for me to spend a sabbatical year visiting his unit.
So I just went about half a mile across town from UCL to this outpost and set myself up there.
And that was interesting.
There were a lot of very clever people there,
I went to all their seminars, and learned quite a lot.
Zoubin Gharamani was there at the time, and I gave talks to them,
but nothing really ever came of it.
We never developed any joint work or anything.
But it was a very interesting time,
and I got to learn a bit more about what was going on in those areas.

\textbf{Vovk:}
I know you have a lot of interest in evolution.
Earlier you said that biology was never a big thing for you.
But evolution, I think, is a big thing for you,
and I thought it might be connected with your college.
You were in the Darwin College, and it's connected to the Darwin family.

\textbf{Dawid:}
That's random happenstance.
It wasn't.
Darwin College post dates Darwin by a century or so.
Darwin went to Christ's, which has his collection of memorabilia and notes and everything.
The only reason Darwin College bears the name is because it's housed in premises
which used to belong to Darwin son, George Darwin.
So it was not named after Charles Darwin, actually it was named after George Darwin.

\textbf{Vovk:}
Also Fisher is a big name in evolution,
and about half of his books are on statistics and another half on evolution.
Is there anything interesting to say about your interest in evolution?

\textbf{Dawid:}
My interest is very superficial, very much dilettante.
I just find it fascinating.
I'm still more interested in population genetics and things like that
than I am in micro genetics, even though that's the big thing.
As I said before,
had I been entering university now,
I might have decided to study genetics,
realizing probably if I'd been lucky, I'd have come away with a third class degree.
Because, you know, like a lot of students, their degree of interest in the subject
isn't always matched by one's ability in it,
and I don't think I have had a great ability in it,
but I do find it fascinating.

\section{Bayesian statistics}

\textbf{Shafer:}
What was the state of Bayesian statistics when you came on the scene?

\textbf{Vovk:}
James Berger in his 2004 conversation [\citet{Wolpert:2004}]
says that when he graduated in 1974,
there was, in his perception, a lot of excellent Bayesian work in the UK:
Dennis Lindley, you, Adrian Smith.
But the US was a desert.
What was your impression?

\textbf{Dawid:}
I've got no impression of the US,
but let me talk about the UK.
There were so few Bayesians around.
I mentioned David McLaren.
Dennis Lindley had been to visit Jimmy Savage.
Savage had already written his book on Foundations of Statistics [\citet{Savage:1954}].
I think he was trying to find foundations for what statisticians already did
and then realized that there were none,
and what he was doing was basically
giving foundations for Bayesian statistics.
And I think Dennis was already independently doing similar things.
He's got an excellent paper in 1953 [\citet{Lindley:1953}],
which is almost Wald's decision theory in a nutshell.
It was just called ``Statistical inference''.
I think it was almost independently done,
but very much in the same tradition as Wald,
with decision functions and admissibility.
Bayesian methods as technical tools for getting admissibility,
but no more than that,
because he does have the memorable sentence there, which is,
``I am, of course, a confirmed frequentist''.
This was Dennis Lindley in that paper.
It's very similar to Savage;
just his own investigations of admissibility and complete class properties
were making it clear that you couldn't really be a confirmed frequentist
without at least acting like a Bayesian.
And then they went and collaborated.
But there was almost no Bayesian stuff out there.
So when we had the first Valencia meeting on Bayesian statistics in 1979,
there was basically every Bayesian in the world there,
and there were no more than 50 of us for sure.\footnote%
  {According to \citet{Bernardo:2009},
  there were 28 invited lectures,
  all followed by invited discussions,
  and no contributed papers.
  The overall number of attendees
  (surely not all of them Bayesians)
  was 93 (from 13 countries).}

\textbf{Shafer:}
In the US, some Bayesians were at Business Schools,
others were in epidemiology or doing actuarial work.
There must have been other streams in Britain too.
I think you got I.~J.~Good working at a secret institution.

\textbf{Dawid:}
That's true.
But the problem was it was all hush hush.
There was Turing, of course, a devout Bayesian essentially, said Good.

\textbf{Shafer:}
And was Harold Jeffreys still around?

\textbf{Dawid:}
Yes, he was.
But Lindley went to Jeffreys's lectures and said they were completely incomprehensible.
There were these giant precursors.
Good, of course, came out of Bletchley Park,
and, because the wonderful things he'd done there were under the Official Secrets Act,
wasn't able to tell you how wonderful he was,
and therefore nobody in Britain wanted to appoint him.
In fact, I do remember, when I was a Diploma student,
the first RSS meeting I went to.
David Kendall came along to the tea room
and said Jack Good is giving this talk at the Royal Statistical Society.
So we went to that.

There were a few wonderful people, but they were very, very few.
And in my time there was Dennis.
Dennis had written this lovely textbook [\citet{Lindley:1965}].
Although he essentially disowned it later
because, he said, he was trying too hard to find Bayesian excuses
for what people were doing anyway.\footnote%
  {See, e.g., the episode described on p.~\pageref{p:disavow}.}
When Dennis came to UCL,
he really wanted to make it into a Bayesian department.
I think he had a little bit of success at that.
At Aberystwyth, he had Mervyn, he had Rodney Brooks,
and he brought them with him to UCL.
But he also had this rump of people left over who weren't very willing to play ball.
So although we had some very good Bayesians,
it was never really a fully Bayesian department.
And whenever Dennis tried to introduce a new Bayesian course into the syllabus,
there were quite a lot of objections [smiles].
So he didn't always get his way.
Nevertheless, it was a good place to be a Bayesian.

Being a Bayesian then was a very different world from what being a Bayesian is now,
because we didn't have computers.
And if we had computers, we didn't have methods to use all the computers.
We had a big mainframe in the far corner of the campus,
but that was it.
So being a Bayesian was doing fairly simple algebra and computation,
and a lot of philosophizing,
because you couldn't really do any proper data analysis.
We didn't have the tools, and that has changed.
Now people do a tremendous amount of wonderful data analysis
using complex computational methods and rather less philosophizing.

\textbf{Vovk:}
A more philosophical question.
Bayesian statistics is not just about using the Bayes theorem,
as it also includes Bayesian modelling.
But what is the role in it of Cournot's principle
(a prespecified event of a small probability
is not expected to happen in a single experiment)?

\textbf{Dawid:}
I don't think Cournot's principle is anything which most Bayesians would recognize.
No, I think there's only one rule of Bayesian statistics,
which is ``all uncertainty is measured by probabilities''.

\textbf{Vovk:}
But you are interested in testing Bayesian models,
so to me it looks like using Cournot's principle:
when you see something unusual,
you may want to revise your model.

\textbf{Dawid:}
Yes, but I'm not a typical Bayesian.
I'm a very untypical Bayesian.
So yes, I have some rather bizarre interests,
but I wouldn't call those Bayesian interests.

\textbf{Shafer:}
So can you tell us more about what you see as typical Bayesian today?

\textbf{Dawid:}
Well, what I see is tremendous focus on really important applications,
very complex modelling.
Most of it requiring extremely deep computation,
which needs very deep computational methods which have been developed.
I remember, many years ago, Adrian Smith and I organized a conference,
in conjunction with what was then called the Institute of Statisticians,
called ``Practical Bayesian Statistics'',
and Maurice Kendall, who was a big name in statistics,
went on record to say that's a contradiction in terms;
there's no such thing as practical Bayesian statistics.
And he was pretty much right because you couldn't actually compute anything.
But things changed.
Now, you could say that practical statistical analysis is almost easier to do
the Bayesian way than any other way.
We have the technology and the computation.

\textbf{Vovk:}
You paid a lot of attention to Bayesian modelling.
Is it fair to say that it is based on intersubjective models;
models shared by, as you said, bevies of Bayesians?

\textbf{Dawid:}
Yes, that was one theme,
where I asked myself the question where do statistical models come from?
Why do we model things in a certain way?
And what are foundational reasons why we can do that?
And so I started thinking
that maybe a model is what is common to different people
who have different views,
what they can discover, something that they share,
and that's the model.

\textbf{Vovk:}
And what about specific models, like the Gaussian model or exchangeability?
The exchangeability model, of course, was already in de Finetti,
but what about later models?
For example, did you know about Martin L\"of's and Steffen Lauritzen's work
when you started thinking about it?

\textbf{Dawid:}
I did start interacting with Steffen Lauritzen at an early stage---%
I must have seen some of his work on extreme point modelling,
something which really interested me,
so I invited him over.
He came as a visitor, and we certainly shared a lot of common interests.
This was, I think, when I was at City University,
so in the late 1970s.
I still think the work on extreme point modelling
is very beautiful.
And as you say, Per Martin-L\"of did the same sort of thing.
It was the Scandinavian School, if you like.

\textbf{Vovk:}
The way of representing it as intersubjective models is really attractive.

\textbf{Dawid:}
I liked it as an idea,
but I don't think anybody else has followed it up in any big way.
Maybe it doesn't even need to be followed up much.
It's just a way of thinking about things.

\textbf{Vovk:}
My impression was that the book by \citet{Bernardo/Smith:2000},
which is perhaps the canonical book on Bayesian statistics,
follows you in that intersubjective aspects are important for them.
They say it explicitly on page 166 \cite[Sect.~4.1.1]{Bernardo/Smith:2000}].

\textbf{Dawid:}
Adrian had an early paper in intersubjective modelling,
which was based on spherical symmetry.
There are two approaches.
One is to do intersubjective modelling via extreme points,
things that Persi Diaconis did with sufficient statistics,
and the other is symmetry structures which people can share.
So it's a question of what do you have in common,
and then how do you build a model out of that.
John Kingman and Adrian worked on spherical symmetry.\footnote%
  {See \citet{Smith:1981} for Smith's paper
  and \citet{Kingman:1972} for Kingman's paper that Philip mentions.
  According to \citet{Kingman:1978},
  Kingman's result was actually due to \citet{Freedman:1963}.
  See \citet[Sect.~11.6.4]{Vovk/etal:2022book} for a short history.}

\textbf{Vovk:}
I think testing Bayesian models is also important for \citet{Bernardo/Smith:2000},
who must be Bayesians.

\textbf{Dawid:}
There's no doubt about their credentials [laughs].

\textbf{Vovk:}
They have chapters on modelling and remodelling,
and the latter has a section on model rejection.
But what about Dennis Lindley?
Was he ever interested in testing Bayesian models?

\textbf{Dawid:}
No, not that I recall.

\textbf{Vovk:}
I find it surprising.
It seems Dennis Lindley was extremely open-minded.
He conceded the validity of the marginalization paradoxes
that you discovered,
and he did it very gracefully and criticized his own book.
What do you remember about it?

\textbf{Dawid:}
Yes.
The record is there in the journal [\citet[pp.~218--219]{Dawid/etal:1973}]
roughly as it happened.
I do remember him getting up, speaking at the meeting,
and basically saying that he was disavowing\label{p:disavow}
his previous attempts to reconcile Bayesian and classical statistics,
because these paradoxes meant that it couldn't be done consistently.
I don't know if he actually changed what he did in any serious way after he said that [laughs].
Well, maybe he did.
Because about that time he was developing ideas
that he worked on with Adrian Smith,
including hierarchical modelling and Bayesian hierarchical models,
which was a move away from what classical statisticians were doing.

\textbf{Vovk:}
So he accepted his fallibility as a person.
If he can sometimes be wrong,
it seems he should accept that his Bayesian model can be wrong.
Maybe it does not necessarily lead you to Cournot's principle,
but do you think Cromwell's rule was his only answer
to the problem of testing Bayesian models?
(You have to assign a positive probability to any eventuality.)
Was there anything else?

\textbf{Dawid:}
Nothing that comes to mind, particularly, no.

\textbf{Shafer:}
When did you start thinking about and working on testing?

\textbf{Dawid:}
The first thing I did on it was the calibration criterion.
``The well-calibrated Bayesian'' [\citet{Dawid:1982}] was my first paper
in that whole area,
and as I recall that came out of a very interesting collaboration.
There was a twice yearly meeting between some very eminent clinicians,
doctors at the Royal College of Physicians,
and some of our statisticians at UCL.
There was a wonderful very eminent physician called Wilfrid Card.%
\footnote{The full name is Wilfrid Ingram Card.
  In 1966--1974 he was a Titular Professor of Medicine at the University of Glasgow.
  In 1974 he retired his position at the University of Glasgow
  but continued as an active researcher
  at the Diagnostic Methodology Research Unit at the Southern General Hospital.
  He was fellow of the Royal College of Physicians
  and the Royal College of Physicians of Edinburgh and Glasgow.
  Died in 1985.}
He had actually worked with Jack Good.
And he was involved in setting up this little group at the Royal College of Physicians.
They were interested in introducing statistical decision-theoretic ideas
into what they called test reduction.
Because of limited resources,
they were thinking about how do we, in a principled way,
use our resources to the best effect.
They called it test reduction,
but essentially it was statistics applied to clinical decision making.
That was fascinating.
We met twice a year for many years,
with very fruitful discussions,
and it veered at one point into probabilistic predictions.
And that's what was at issue there.
And then I started thinking:
What are these?
And when can you make sense of them?
When do they have any relationship to what you want to know?
And so I started thinking about testing probabilities.
That became ``The well-calibrated Bayesian''.
So it came out of that very applied context.

\textbf{Shafer:}
So that was during the early 1980s then.
``The well-calibrated Bayesian'' was 1982.
But you said the seminar had gone for many years.

\textbf{Dawid:}
It probably started in the early 1970s.
I was fascinated by it all and was quite involved in it.
Eventually we discussed a lot of interesting things in the early stage.
Then the personnel changed, people dropped out and people dropped in,
and then after a certain point I realized that we were just going full circle,
discussing again the things that I thought we'd sorted out ten years earlier [laughs].
So I dropped out.

\textbf{Vovk:}
In your position statement
(2004 paper in \emph{Statistical Science} [\citet{Dawid:2004}])
you write
``We regard a probabilistic theory as falsified
if it assigns probability unity to some prespecified theoretical event $A$,
and observation shows that the physical counterpart of the event $A$
is in fact false.''
It's about probability one rather than probability close to one
and so seems impractical.
Can you comment on this?

\textbf{Dawid:}
It's completely unachievable.
It was exactly probability one,
because there's a nice theory about it.
There's a nice theory about probability close to one as well,
but I wasn't aware of this [laughs].
That was Cournot's principle, as you know.
Or one version of it.

\textbf{Vovk:}
I think Cournot was talking about probability zero.
Glenn, is it right?

\textbf{Shafer:}
Well, that's interesting.
He didn't talk about this too much, but it's true.
In his 1843 book, he said infinitely small.
And then in 1851 he had another book
where he mentions this infinitely small several times.
But finally in 1875, he says infinitely small does not really mean zero.
It just means very small [laughs].
I don't want to put words in Philip's mouth,
but maybe he's old fashioned, and when he says infinitely small,
he doesn't really mean it.

\textbf{Dawid:}
Well, I really didn't mean probability one [smiles].

\textbf{Shafer:}
But of course Cournot came before the strong law and all that.

\textbf{Dawid:}
So in those days they didn't even have events of probability one to play with.

\textbf{Shafer:}
Not really.
Yes and no.
Ampere had an event of probability one.
He proved that the gambler will be ruined with probably one.
But mostly it wasn't like that.
They were using Bernoulli's theorem for most of what they were thinking about.

\textbf{Vovk:}
In one of your versions of Cournot's principle
you replace probability unity by a gambling strategy.
For example, in your \emph{Annals of Statistics} paper [\citet{Dawid:1985}].
And instead of saying a strategy should be specified in advance,
you said it should be computable,
which is a valid replacement when we are talking about infinite sequences.
Is there anything interesting to say about the role of computability in statistics in general?

\textbf{Dawid:}
I think the answer must be yes, but I think that story probably still has to be told.
My own dabbling was very much on the fringes.
I was trying to find some way of getting a countable number of things to think about and worry about,
in order that the intersection of a countable number of events of probability one would still have probability one.
Uncountable doesn't work.
And I wanted to have a criterion which says
that if something to which you assign probability one doesn't happen, then you're in trouble.
And so I wasn't really particularly concerned with computability as such,
more with a very broad-ranging principle which I could use,
as broad-ranging as it could be and providing me with just a countable number of things to think about.

\textbf{Vovk:}
Maybe something expressible by a formula of set theory?

\textbf{Dawid:}
Yes, there could be different ways of saying it.

\begin{figure*}
\begin{center}
  \includegraphics[width=10cm,trim={17.5cm 4cm 1mm 1mm},clip]{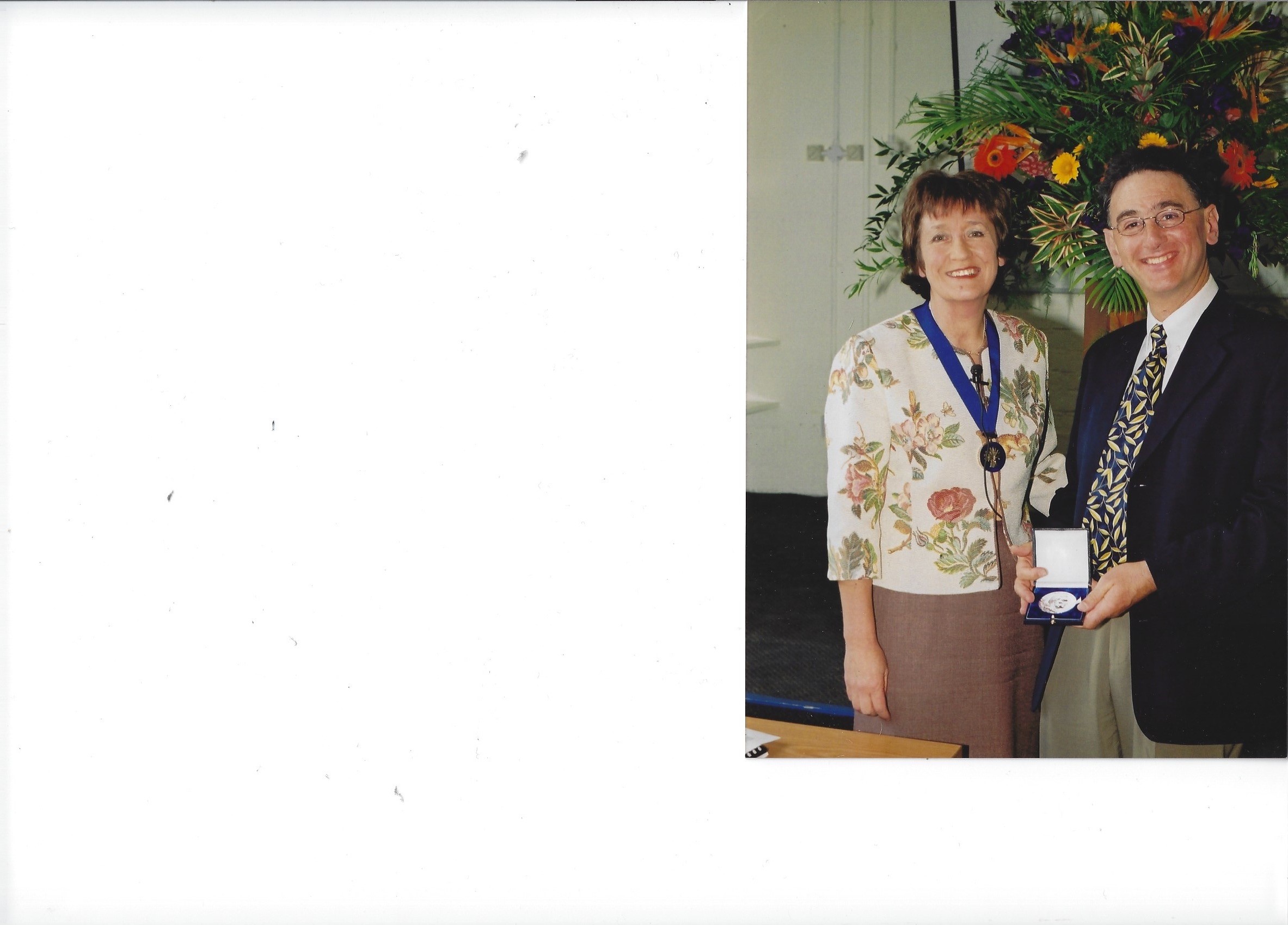}
\end{center} 
\caption{Receiving his Guy Medal in Silver from RSS President Denise Lievesley in 2001}
\end{figure*}

\section{Fiducial and frequentist statistics}

\textbf{Vovk:}
You have a long-standing interest in Fisher's views,
which were a focus of your 1991 RSS discussion paper [\citet{Dawid:1991}].
How would you summarize now the difference
between Fisher's and Neyman's views of statistics?

\textbf{Dawid:}
They had really very different views.
Neyman essentially was tied to this repeated sampling
or parallel universe idea of interpretation of probabilities
and decision making.
Fisher was also doing, as it were, sampling-theoretic computations,
except when he was doing fiducial stuff
(and even there, there was a connection of course).
But Fisher made this big distinction
between what I call the production model,
the actual thing that would produce the data you've got,
and the inferential model,
which is appropriate for analysing the data after you've got it,
and in so many different examples of things he did,
these were clearly very distinct.

\textbf{Vovk:}
Can we say that Fisher was midway
between Neyman's and the Bayesian points of view?

\textbf{Dawid:}
I don't think any of Neyman, Fisher, or Bayes would agree with this statement [laughs]!
I think most statisticians now who haven't done a lot of research into it
will probably not realize how big were the differences
between Fisher on the one hand and the Neyman--Pearson approach on the other.
Because they're all classical statistics,
and various specific tools will be borrowed,
some from Fisher and some from Neyman and Pearson.
There's a ragbag of things.
So the fact that they really had very different outlooks on what to do
probably is lost on most modern statisticians.

And with Fisher, we come to fiducial.
What was he looking for?
Was he looking for probabilistic statements about parameters?
Or was there some other motivation?
I don't know.
He was very anti-Bayesian,
but it seems that by developing fiducial theory at least,
he had some of the same instincts as a Bayesian of wanting
to quantify all uncertainty in terms of probability.
Even uncertainty about unknown parameters.

\textbf{Shafer:}
Could you give an example where Fisher is making this distinction
between production and inferential models?

\textbf{Dawid:}
I've got one or two specific examples in that paper.
There's a very good one about a contingency table
which has been developed with some sort of optional stopping process
whereby the data have been gathered.
But he says explicitly that it should be analyzed
as if it was just fixed sample size in advance.
Which it wasn't, and even conditioning doesn't do that for you.
So that's a different inferential model from the production model.

\textbf{Vovk:}
Let me ask about an approach to modelling
which, I think, is standard in frequentist statistics
and is very different from intersubjective modelling.
Many statisticians talk about real parameters;
at least in their minds parameters are real.
Do you think it's something that has a non-vacuous scope,
something that makes sense?
For example, you can have some physical theory
that depends on a parameter such as the speed of light.
It will be measured eventually, but at this point in time,
you don't know it, and maybe some probabilities depend on it.
What do you think about this approach to modelling?

\textbf{Dawid:}
You wouldn't normally interpret the variance of a distribution
as being some real measurable quantity.
It's just a description of a distribution.

There are extrinsic and intrinsic parameters.
So extrinsic are things like the speed of light, distance to the sun,
or whatever things which you might measure with error or whatever.
But they have an external meaning outside your model,
and they can apply to different models.
And intrinsic parameters are other things which are created,
maybe like de Finetti creates probability of success out of exchangeability.
They don't correspond to any specific quantity that you would go out and measure;
that is just part of the model, like a variance.

But there's a difference of interpretation,
so probably a lot of statisticians will say
the bias of a penny is an extrinsic parameter;
it's really there.
But if you take de Finetti's approach, it's intrinsic,
it's just something which describes the model.

\textbf{Vovk:}
Would it be fair to say that for you extrinsic parameters exist,
but they don't play a big role?

\textbf{Dawid:}
Yes, I think that's fair.
I mean, they haven't played an important role in what I've done.
That's true.
But that's not to say they're not important.

\textbf{Shafer:}
We talk about so many people being Bayesian,
but a lot of people, as I understand them,
are using Bayes as an exploratory tool.
But then they want to end up with something that has frequency properties,
as they might say.
Is that your perception?
Is that a lot of the Bayesians are not quite Bayesian
the way you would want them to be?

\textbf{Dawid:}
There's of course the whole enterprise of so called objective,
or what I call objectionable, Bayesian inference [laughs].
It's much, much bigger than subjective Bayesian inference,
even though it's provably internally inconsistent,\footnote%
  {As demonstrated by marginalization paradoxes.}
but never mind.
And there is also this desire,
which of course Lindley had in his textbook,
to somehow find parallels between Bayesian answers and frequentist answers.
I remember a conference in which Jos\'e Bernardo was talking
about, I think, a prior for the binomial,
but the sole purpose of what he was doing was to reproduce classical answers.
I thought, if we're going to do that, why not be a classical statistician?

\textbf{Shafer:}
What is your perception of what goes on both inside stat departments,
but also outside stat departments?
In your perception,
what fraction of the statistical world is Bayesian in both cases, and how?
How much of it is sort of Bayesian for real [laughs]?

\textbf{Dawid:}
Alright, I have got to consult my pet psychologist,
who will then interrogate me in my opinions about these things,
and I come up with distributions for these probabilities [laughs],
because they're pretty vague priors.
Obviously,
there's a lot of Bayesians around now as compared with when I was young.
And a lot of them, I would imagine,
are really concerned not so much with the methodology
as with the application to resolving real problems,
and often they will choose convenience priors to do that.
They're using the Bayesian machine.
But there's always the question of what are the inputs to the machine,
and they maybe not thought about it a lot, and maybe it doesn't matter.
Maybe sometimes it does.
I think there's a pretty tiny rump of people
who really care about assessing prior probabilities.
Getting them right and knowing that it matters.
So the real subjectivists are very few in number, I think.

\textbf{Vovk:}
Do you have any feeling of whether we are moving
towards more or less agreement in statistics?
Is it becoming more fragmented?
Or maybe things like BFF (Bayesian, Fiducial, Frequentist) workshops
are achieving their goal,
and there will be some unified point of view.

\textbf{Dawid:}
I don't think there will ever be a unified point of view,
but it's certainly less fractious than it used to be.
When I was young, there was real animosity between Bayesians and non-Bayesians.
Some people, such as Lindley, suffered a lot from just being completely dismissed,
pooh-poohed.
People weren't willing to listen to him,
because he wasn't one of them [smiles].
He was almost the only one.
There were a lot of serious disagreements between the different camps,
but usually not argued in any rational way.
And I think that died down a lot.
It doesn't mean that the issues were resolved at all,
but it's just people are more prepared to live and let live.
And as you say,
the BFF meetings are excellent in trying to bring people together
and discover common ground.

\section{Prediction and prequential statistics}

\textbf{Vovk:}
Prediction,
which is the topic of this special issue of \emph{Statistical Science},
is usually regarded as only one concern for statistics.
How do you see the role of prediction in statistics?

\textbf{Dawid:}
I think prediction is a special case of something more general,
which is concentrating on observables.
There is a lovely phrase Dennis Lindley used to have.
He said, ``We should be concentrating not on Greek letters
but on the Roman letters.''
Not the $\theta$s, but the $x$s and $y$s.
I think I was always primed for this,
but de Finetti was a big influence on me;
thinking about the importance of concentrating on observables
is one of his big things.
Actually, Dennis Lindley in our early days in the department
invited him over.
And de Finetti gave a series of lectures
which were very interesting and informative and influential on me.
So I got this general feeling:
what's the point of making inferences about an unknown mean
where we'll never know if we got it right or wrong?
Let's put our head above the parapet and make statements
which we can actually be tested on.
And that means we have to talk about observables.

\textbf{Vovk:}
Prediction is the basis of prequential statistics,
your approach to statistics.
Did the idea of it come from meteorology,
or did it have other sources?

\textbf{Dawid:}
There are two aspects to it.
One is the assessment of probability forecasts,
whether or not they're sequential.
These are things to do with proper scoring rules,
which were wonderfully developed in the meteorological literature.
They did some fabulous work.
And then there are things to do with the sequential nature of things,
which also came from there, with calibration methods for example.
Alan Murphy was a very big contributor to that sort of thing,
as in his work with Winkler.
So yes, I was very influenced by the great ideas which came from meteorology.
And which really were much more sensible, I thought,
than the sort of things that people in statistics were saying and doing.
I think that probably came after I wrote my calibration paper
(``The well-calibrated Bayesian'' [\citet{Dawid:1982}]).
I've given you some background on that.
I use weather forecasting as an example there,
but I think I probably followed the literature up after that.
So it was ``The well-calibrated Bayesian''
which made me think seriously about the sequential nature of things.
And that morphed into the general prequential approach.

\textbf{Vovk:}
Did you ever meet Seymour Geisser?

\textbf{Dawid:}
Oh yes. Seymour was a very fun guy.
There were a few people who were doing Bayesian predictive things.
Seymour Geisser was one.
Also Aitchison and Dunsmore.
There was an excellent book by them
about statistical prediction analysis,
which I found very nice.
And Raiffa and Schlaifer worked out predictive distributions.
There were issues about computing predictive distributions and things like that.
So there was, in a sense, an emphasis on doing prediction,
but I had an even stronger emphasis on it:
really that was all you should be doing.

\textbf{Vovk:}
What's the current status of the prequential principle,
and do you regard its different varieties as equally important?

\textbf{Dawid:}
Oh well, I think I'm as confused now about it
as I was when I introduced it,
as it's a rather vague principle.
It's got, as you know, various flavours, weak and strong.
So it's a meta-principle more than a principle.
Is it having an impact on anything anybody does though?
I don't think it is really.
Except the two of you, of course.
You are about the only people, and maybe a few others,
who've taken it very seriously and run with it, which is lovely.
And of course it's got these links
to all sorts of non-probabilistic, game-theoretic approaches,
where really you can't do anything else;
the prequential principle isn't a principle, because you can't avoid it.
It was almost built into the way you do things;
that's all you can do.
Just do comparison of outcomes with forecasts sequentially,
and there is nothing bigger to do the comparison with.
Whereas with a statistical model underlying it,
you've always got this option of thinking
about repeated trials or something else,
and different ways of thinking about it.
And it was just excess baggage.
So throw away excess baggage.

The work you and I did, Vladimir,
was also interesting here [\citet{Dawid/Vovk:1999}].
I think it is showing how you can get a lot of technical support
for the prequential principle
that really seems to deliver the goods, most of the time anyway.
It's an aspect of what I said about Fisher,
which is the difference between the production model and the inferential model,
so there may be some complicated stochastic process producing your forecasts and data,
but you don't care about it,
your inferential model is just based on the forecasts you actually got.
Maybe you could have thought of them as arising independently, it wouldn't matter.

\textbf{Vovk:}
What's the current status of prequential models?
Are there any interesting applications of them?

\textbf{Dawid:}
Cox's partial likelihood is a sort of application.
There's the log-rank test,
where you have correlated contingency tables in time,
and you combine them as if they were independent.
People have been doing these things without defining them.

\textbf{Shafer:}
Let me interject a question about Kolmogorov--Doob.
Was that important in your education and world?
Obviously the prequential model is not working within that framework,
but was statistics within that framework when you came up?
Did you perceive it as having any importance
to the way you and others were thinking about statistical inference?

\textbf{Dawid:}
There was no alternative.
It was the atmosphere in which we lived.
What else could you do?
Had to breathe it.

\textbf{Shafer:}
Is that right?
Did you have to study it as part of your training to be a statistician?
Or was it just lip service?
There's that over there, somebody told us.

\textbf{Dawid:}
Well, we had courses in probability theory,
and then when you come to doing statistics,
there're some results in probability theory that you apply,
such as central limit theorems.
For the asymptotic distribution of the maximum likelihood estimator,
you need various limit theorems.
So in that sense, you're using results from probability theory,
standard probability theory.
There wasn't anything else you could do.

\textbf{Shafer:}
But you weren't using filtrations, were you?

\textbf{Dawid:}
Oh well, I wasn't into that in a big way,
but I think that, in a sense, it's very nice stuff.
You've got to work with sequential models,
stochastic processes and things.
You know, I think that all that Doob stuff is very beautiful,
it's got big ramifications, and there's a lot of it out there,
even if maybe there's more of it out there than there needs to be,
because we have less structured way of doing those things with game theory.

\textbf{Shafer:}
Thank you.
I've been, you know, interested in at what point Kolmogorov--Doob became important.
Certainly it wasn't until after World War Two.
But I guess by the time the two of us were in graduate school,
it was taken for granted.

\textbf{Vovk:}
You mentioned the course about axioms,
or at least that started from the axioms of probability.
So it's like Kolmogorov.
But probably it was just Kolmogorov, not Doob.

\textbf{Dawid:}
I don't think I came across Doob until much later.
Actually,
I seem to remember David Kendall in Cambridge
may have given a course on martingales.
Possibly.
But it didn't make a big change to what I did;
that was much, much later on.
When I was working on ``The well-calibrated Bayesian'',
I realized that I needed some martingale theory.
I picked up what I needed, which wasn't very much.

\textbf{Shafer:}
So how did you discover Jean Ville?

\textbf{Dawid:}
That's a good question,
because he wasn't very well known, was he?

\textbf{Shafer:}
No.

\textbf{Dawid:}
I pass, can't remember [laughs].

\textbf{Vovk:}
You had a remark about him in the 1985 paper [\citet{Dawid:1985}],
but to me it looked like an afterthought.
You were doing von Mises in most of the paper,
but then suddenly you have this beautiful remark
saying that you can do it like Ville.

\textbf{Dawid:}
Well, I must have discovered him by then.

\textbf{Vovk:}
On the other hand,
while I thought von Mises's approach was very restrictive in general,
in your paper it looked exactly what you needed
to discuss calibration
and then to prove Jeffreys's law
(well-calibrated forecasters will agree eventually).
For that von Mises's definition was enough.
You didn't really need Ville.

\textbf{Dawid:}
I'm not sure that von Mises was very influential in that,
as I recall.
One thing I discovered at the same time,
and I'm still fascinated by the connection,
is Blackwell--Dubins.
The result where you have
two mutually absolutely continuous distributions
and you keep doing sequential updating.
You'll end up with probabilities that converge almost surely,
which is like emergence of objective probabilities.

\textbf{Vovk:}
And it's not just for one step ahead,
it's for infinitely many steps ahead.

\textbf{Dawid:}
It's for the infinite future, yes.
Convergence in a suitable metric.

\textbf{Vovk:}
But in your 1985 paper you talked a lot
about subsequence selection rules, so it's like von Mises to me.

\textbf{Dawid:}
Yes, I see what you're getting at.
Yes, absolutely.
It was under suitable subsequences.

\begin{figure*}
\begin{center}
  \includegraphics[height=8cm]{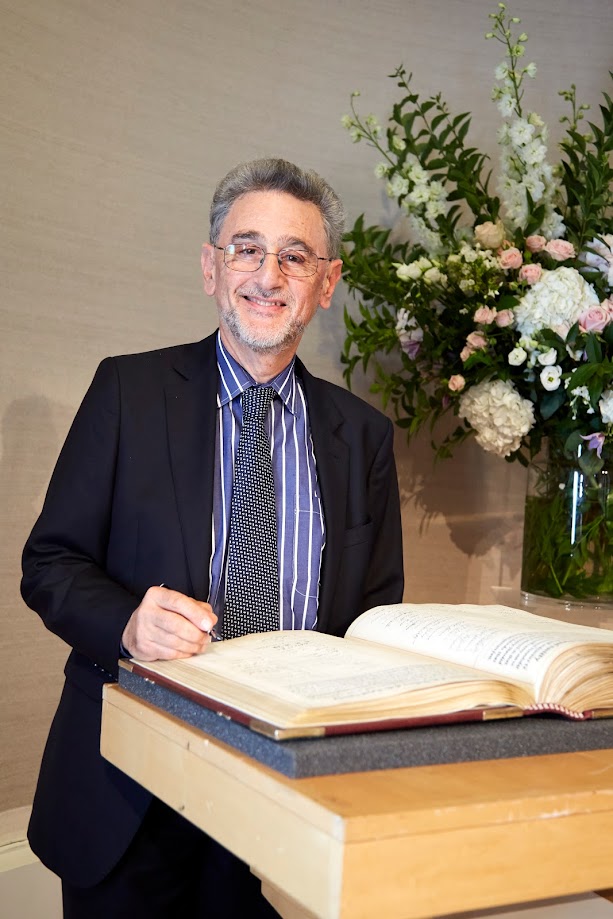}
\end{center}
\caption{Admission as Fellow of the Royal Society, London, 13 July 2019}
\end{figure*}

\textbf{Vovk:}
I know the word prequential consists of ``predictive'' and ``sequential'',
but it seems to be about one-step-ahead predictions.
How important is the one-step-ahead aspect, if at all?

\textbf{Dawid:}
That's a good question.
It was important to the way I developed things.
To drop it changes the theory a lot.
To get nice theory and great things coming out of it,
it seems to require that it be one step ahead.
And you get a link with your sequential games and all the rest of it.
If every day you forecast a week ahead,
you don't seem to get the same sort of lovely things coming out;
things start depending on underlying assumptions about stochastic model
which otherwise they wouldn't.
I'm not tied to that as being necessarily the right thing to concentrate on,
but it's the only thing I was able to work with,
and it does seem to have some beautiful properties.
Predicting a week ahead is a pretty important problem,
but I don't know what to do with it.

\textbf{Shafer:}
Our 2001 and 2019 books also concentrate on one-step-ahead prediction.
They were strongly influenced by your work on prequential methods,
but do you regard the approach of those books as fully prequential?

\textbf{Dawid:}
I think your work is fabulous.
My critical appraisal will be 100\% positive, and it's essentially completely prequential.
When I was developing the ideas, nobody really seemed to be interested.
I gave my 1984 paper [\citet{Dawid:1984}] at the Royal Statistical Society,
there were a few other things I followed up with, but nobody else cared.
I could foresee quite a lot of important problems which ought to be tackled,
but realized I didn't have the mathematical skills to do.
And so how delighted was I when people like the two of you
come along and show that they can tackle these things
and have the skills and do wonderful things?

\textbf{Shafer:}
I'll ask a small question.
Back when I was interested in your 1984 paper [\citet{Dawid:1984}],
I remember talking it up to some colleagues whose reaction was:
``How can this be fundamental?
Doesn't most of statistics look at the observations as batches?''
Do you think that has changed in statistics?
Does sequential observation now seem more fundamental than it did 40 years ago?

\textbf{Dawid:}
If you remember, there was a story of economists saying that probability theory doesn't apply to economics
because things across time aren't independent identically distributed.
For so long, and it's still true of 97\% of everything done in statistics and machine learning and everything,
the fundamental assumption is basically we have just a bag of exchangeable goodies.
And I thought that was just so limiting; how boring.
There's a big wide world beyond that.
And one aspect of that world is sequences (not the only thing)
where you may have almost completely arbitrary structure
as a lot of the prequential stuff allows.

\textbf{Vovk:}
Let's talk about prediction more generally.
We know that you have a very interesting picture of three levels of prediction\label{p:three},
according to the assumptions made.

\textbf{Dawid:}
Yes, there are three universes you can work in,
and it's interesting to understand relationships between them.
One universe is where you have a statistical model,
and you believe the model,
and you do computations under some value of the parameter
in that model.
And you see what happens,
that's the classical statistical thing.

And then, moving on a slightly more general stage,
you use a statistical model as a tool,
and you do computations with it,
but you don't actually have to believe
that the data are generated from something in the model.
This is to do with the prequential principle,
where in a sense you aren't really worried so much
about whether you got the model right or not,
but just what you actually got to see.

And finally, you throw away probability models altogether,
and do the learning with expert advice,
or game-theoretic probability, or whatever.
Then you do worst case analysis.
And there are fascinating connections between those three different universes.
So that's something I'd like to understand better than I do.

\textbf{Vovk:}
Can you elaborate on that?

\textbf{Dawid:}
In classical statistics,
the model may be parametric or something,
but you act as if you know you are right.
You may not know which member of the model is right,
but you are sure one of them is,
and therefore all you are concerned about is behaviour under a typical member of the assumed model.

The next stage is where you have a working model.
It's still a model, but you're not convinced that you got it right,
and therefore you think of procedures which are based on the idea that the model is correct.
That gives you your procedures, but your evaluation does not assume that the model is correct.

And the final thing is when you get to prediction with expert advice,
where you don't even assume any distributions at all,
and you are just considering worst case aspects.
Also the connections between those are quite interesting.
They are all connected, but they are obviously very distinct as well.

\textbf{Vovk:}
We know that you have contributed to all three of those areas.
Can you say that one of them has dominated your research?

\textbf{Dawid:}
First of all, as a convinced subjectivist,
I agree with de Finetti that probability does not exist,
and so there is no such thing as the true model.
Much of statistics works with and talks about the data generating process.
I don't believe there is a data generating process.
I think there's just data.
And we have models, and our models have probability,
but Nature doesn't have a probability to generate data.
We have models to understand data.
If you believe there is a data generating process,
then you are in the first two universes.
Either I've got a model which includes it,
or I've got an interesting model to help me to do something:
there is a data generating process, but it's not there.
My fundamental philosophical attitude is that
there never was a data generating process in the first place.
So (a) it can't be in your model, and (b) it can't even be outside your model.

Mathematically, as we developed in our joint work [\citet{Dawid/Vovk:1999}],
Vladimir,
there's a lot of common ground between the second and third universes.
Take the idea of what happens when you assume there's a process,
but it's not in your model, and it could be arbitrary.
It's almost a worst case analysis again,
as it's going to work against pretty much any data generating process.
In a recent paper with Ambuj Tewari [\citet{Dawid/Tewari:2022}]
we show that statistical learnability of general non-iid stochastic processes
is equivalent to worst-case online learnability.

And then the step beyond that is that your method works against any sequence of data,
as in prediction with expert advice,
which, I think, is the right philosophical thing to do.
But I'm very out of kilter with most of my statistical colleagues here.
So much there assumes that the job of statistics
is to identify the data generating process.

\textbf{Shafer:}
This term, ``data generating process'', which I share your aversion to,
I think, was pretty much invented in our lifetime, during our careers even.

\textbf{Dawid:}
That particular form of words seems to be quite recent.

\textbf{Shafer:}
Do you have any insight into how it got a grip on our profession?

\textbf{Dawid:}
Well, I think the concept was there long before the three word term.
That was basically in most of the statistics I learned and grew up with,
the idea that Nature had a probability model that was generating data.

\textbf{Shafer:}
This is a historical question,
but I wonder whether it really was there,
or maybe it was there, but people were embarrassed to say such a thing.

\textbf{Dawid:}
If you go back to the early days of statistics,
it really was data centred,
it was histograms, and means, and just data summarization.
The idea of some underlying thing generating the data didn't occur till much later.
Maybe with Fisher or with Karl Pearson,
I don't know.
Pearson had frequency curves.
It would be an interesting historical study.
So this is all speculation,
but I think people,
like Pearson in developing his frequency curves,
were just trying to describe data.

\textbf{Shafer:}
Well, they had populations.

\textbf{Dawid:}
They had populations, and they wanted to describe them.

\textbf{Shafer:}
The generating process was just the sampling process.

\textbf{Dawid:}
Certainly.
I don't know when it actually came in,
this idea of a statistical model, and the model as a collection of distributions,
one of which you think, or at least hope,
is actually what nature is doing somehow.

\textbf{Shafer:}
That history needs written, yes.

\textbf{Dawid:}
Anyway, I don't really have any great insights into that except that I don't like it [laughs].

\textbf{Shafer:}
Our next question is about statistical education.
Should our elementary statistical courses have more about forecasting,
or is that just a different subject than elementary statistics?

\textbf{Dawid:}
That's a good question.
Elementary statistics, what is it?
It's still classification of data, isn't it?
And calculating the means and modes and all the rest of it.
You go beyond that, and you have normal distributions,
and then you have samples from populations.
So it is a bag of things.
Maybe all that stuff has to be the foundation
before you introduce more general stochastic processes or sequential ideas,
which are obviously much more subtle and complicated.
I think they should certainly be emphasized more strongly,
but maybe in a second course.

\textbf{Shafer:}
A different way of asking my question
is whether you see a desirable future
that would have more of a fusion between machine learning and statistics
at an introductory level.

\textbf{Dawid:}
I think, at an introductory graduate level that will be perfect.
Not for the undergraduate syllabus.
It might be an interesting experiment.
Talking about experiments, I remember when Lindley first came to University College,
he decided he was going to teach the first year undergraduates everything
from a subjective Bayesian point of view.
He discovered how hard that was.
The first lesson I think was about exchangeability [laughs].
That's fresh-faced young undergraduate, 17 years old.
A good try, but it didn't last long.

\section{Causality}

\textbf{Vovk:}
Causality is another of your major research interests.
When did you become interested in it?

\textbf{Dawid:}
There was an early paper by Pratt and Schlaifer,
which I commented on [see \citet{Dawid:1984comment}],
but my interest, I think, goes back to the time in the early 1970s when I was at UCL
and Don Rubin came through to give a seminar.
He hadn't yet published his stuff on potential responses,
but he'd done a lot of work on it, and he gave a seminar on it.
And I remember sitting there, listening to Don talk,
and thinking that this is absolutely the wrong way to go about things [smiles].
And he and I have been at loggerheads over that for about 50 years.
I didn't follow that up at the time very much.
This sort of stayed somewhere tucked into my brain.
But I started working on it seriously when Judea Pearl came along.
I got very interested in Pearl
because of our common interest in probabilistic graphical models
rather than causal models.
I was very interested in his stuff there.
And then he moved into the causal side of things.

\textbf{Shafer:}
You have been a prominent voice discouraging talk about counterfactuals in statistics.
Do you regard facts about the future as facts?
If you say something is contrary to what happens that isn't determined yet,
what is your feeling about that?
Do you regard the future as determined?

\textbf{Dawid:}
No, I don't regard the future as determined,
and particularly when it may be affected by what I do.
When I think about facts, I think about, essentially, things that have happened,
things that are the case now.
In a philosophy publication [\citet{Dawid:2007philosophy}],
I made the distinction between hypotheticals and counterfactuals.
Hypotheticals were about the future, but they were hypothesized on something I might do.
Still uncertain.
So I don't think it matters whether you call them facts or no.
I don't think I would want to.
But the term counterfactual refers to consideration of something
which is in contradiction to a known fact,
and it presumably isn't a known fact unless it's already happened.

\textbf{Shafer:}
Do you think your colleagues in statistics are restricting it in that way?

\textbf{Dawid:}
I think there's a lot of confusion over the use of counterfactuals,
which annoys me enormously.
Even philosophers rarely make the distinction, I think a very important one,
between hypotheticals and counterfactuals,
and often the word counterfactual is misused to mean hypothetical.
So, for example, I went to a conference just before lockdown
called ``Counterfactual prediction''.
What it meant was this:
I've got a lot of data, and now I want to use my data
to help me decide what would happen if I did something.

\textbf{Shafer:}
You have made the distinction between causes of effects and effects of causes.
When you talk about effects of causes,
could that be replaced entirely by talk about forecasting, or is there something more going on?

\textbf{Dawid:}
It's forecasting, but hypothetical forecasting.
So if I were to do this,
what forecast would happen?
So it's conditional or hypothetical prediction.
In fact, there is a recent exchange with Judea Pearl
[see \citet{Pearl:2022} and also \citet{Dawid:2022Pearl}]
where he actually complains that my approach is just statistical prediction.
It is [laughs]!

\textbf{Shafer:}
And the business about causes of effects.
I want to interpret that as meaning that you're looking for responsibility.
Is that what it's about, or is there more to it?

\textbf{Dawid:}
No, not really much more than that.
To me that is still Terra Incognita.
I think nobody really has a good handle on what to do about that problem.
Because I still need some sort of counterfactual construction to think about that,
which I don't like.
But I don't have an alternative.
I certainly think that most people don't even make
the distinction between causes of effects and effects of causes,
they just use something like Pearl's structural causal model,
which can be applied in both cases.
But whereas it works fine for effects of causes,
when dealing with causes of effects
there are too many ambiguities.
In my 2000 paper [\citet{Dawid:2000}] I pointed that out.
In that paper I also developed a slightly different, more stochastic, version.
But I haven't been able to escape counterfactuals, which annoys me.

\textbf{Vovk:}
How important do you think are those distinctions between you and Rubin, for example?
Are they just philosophical,
or do they show in practice in the way people do causal inference?
Can they mislead people?

\textbf{Dawid:}
I think they are important,
because I think there's a lot of rubbish out there.
There are problems which you can tackle from either point of view,
with or without counterfactuals,
and get essentially identical answers but with different interpretations.
But there are also problems where, I think,
Rubin's approach gives you meaningless answers.
And the same with Pearl.
You think you've got an answer,
but really you're just getting out some arbitrary assumptions you put in.
Because various things are not identifiable,
but you pretend that they are;
they only become identifiable because you've invented something.

And also I'm a minimalist.
I like economy of thought,
and I don't like introducing extra things you don't need,
like what would have happened to this patient
if he'd been treated differently,
which isn't at all relevant to what's going to happen to a future patient.
You don't need it.
So why complicate things?
Why just mess everything up?
Even when you get the same answer,
I think you're making life difficult for yourself.
I find it actually quite difficult to understand
why people like that approach so much,
because I find it so obviously nonsensical [laughs].

\textbf{Vovk:}
Research on causality has become so important lately.
Where do you think it's heading?

\textbf{Dawid:}
Well, I think there are some dangerous directions.
I'm currently working on a response to some recent ideas of Pearl.
And he's talking about personalized medicine,
and how we should choose who to treat.
He does that using counterfactual analysis.
And I think he's basically completely wrong.
I think it's utterly meaningless,
and he's making a really big play on this
and trying to sell it as a way to do personalized medicine,
which is obviously a big buzzword.

There are actually some rather worrying directions for causality,
but the trouble is that most people aren't very critical of these things,
unlike me,
and these meaningless, dangerous things could very well end up
getting well established.
Another possible danger is the notion that we need a counterfactual understanding of algorithmic fairness,
bringing it into the realm of causality.
There's a lot of good stuff going on, a lot of good stuff will continue,
with lots of great applications.
But to the extent they're going to be new directions,
I am a little worried about them.
Seems like there could be dangerous directions.

\section{Conditional independence}

\textbf{Vovk:}
You have done fascinating work in conditional independence.
It has become extremely popular, everybody is using your notation,
and we know there is more to it than just notation.

\textbf{Dawid:}
Let me just talk about that.
I have a meta-theory that notation is the most important thing in mathematics.
Notation and representation are much more important than theorems and proofs.
For example, graphical representation of problems
instead of complicated algebraic development
where it's hard to see what line followed from what.
You look at a picture, and it hits you between the eyes,
and it's the same with notation.
You get the picture right, you get the notation right, and you don't really need to do anything more.
So I think one of my biggest contributions to theoretical statistics is the introduction of symbolism,
of the notation for conditional independence.
I also had some theory of it, but nobody seems to take that seriously, alas.

\textbf{Vovk:}
How would you describe the place of conditional independence in your research
and in statistics in general?

\textbf{Dawid:}
It has been a unifying thread throughout pretty much everything I've done.
And I've used it in many ways, both very theoretical and quite applied.
For example in forensic inference.
The notation has certainly spread abroad rather well,
although rarely with an appreciation of where it comes from.
But never mind.
I was disappointed
that people still do very clumsy operations with conditional independence
rather than using the simple axioms which I developed in my 1979 paper
[\citet{Dawid:1979}],
which, of course, were essentially the same as what Pearl developed when he went into graphical modelling.
But of course they're not just concerned with graphical modelling.
I think, outside of doing inference on graphs,
people using conditional independence generally make heavy weather of it,
whereas if only they would read the rest of my paper rather than just absorb the notation,
they might have an easier way.
So I think it is, rather surprisingly, an incredibly useful tool across so many different areas.

\section{Administrative work}

\textbf{Vovk:}
You have served the statistical community in lots of different ways.
For the Royal Statistical Society,
you were active in the Research Section as Chairman and Honorary Secretary,
and you were the society's Vice President.
Is there anything else?
Are there any particular achievements you would like to share with us?

\textbf{Dawid:}
I think there's something which probably slipped under your radar,
which is much more recent:
I was a member of what was originally a working group
and then became a fully fledged section of the Royal Statistical Society
on Forensic Statistics.\footnote%
  {In 2013--2015 Philip was a member of the RSS Statistics and Law Working Group,
  and in 2015--2019 he was a member of the Statistics and Law Section Committee.}
I was very involved in that.
We put on a number of very good meetings,
and we made contributions to various guides and things for practitioners.
I think that is one of the most useful things that I did.

When I was Chairman of the Research Section,
that was quite a lot of work,
but it was very interesting.
It was basically to do with assessing submitted papers.
Are they suitable for reading to the society in meetings organized by the Research Section?
So it's a sort of refereeing and editorial job,
which is important and interesting,
and I enjoy it.
It's one of the standard things that we academics do anyway,
at a slightly different level, but not very dissimilar.

I think the forensic statistics work was unlike
most of the other sections of the Royal Statistical Society,
which exist largely to put on meetings.
We did do that,
but a lot of our work was involved in making representations to the Law Society
and bringing out documents to assist lawyers and forensic scientists and things like that,
to help understand the use of statistics in the law.
So it was a bit different from the usual run of things,
and I found it personally very satisfying, very interesting.

\textbf{Vovk:}
For 10 years (1983--1993) you were Head of Department at UCL.
Was it a big drain on your time?
Did you have any time for your research when you were doing that?

\textbf{Dawid:}
I was not a good administrator.
I've never enjoyed that kind of thing,
and I'm certainly not very good at it.
When I had to do it, it probably did eat into my time rather a lot,
I have to say.
It wasn't too bad, and we weren't overwhelmed with boring meetings and things like that,
but we had our share of them.
The department was, on the whole, running on a fairly even keel.

\textbf{Shafer:}
Do you regret taking any of these administrative assignments?

\textbf{Dawid:}
Do I regret it?
I was fine at UCL.
I didn't enjoy administration in Cambridge very much,
and that was largely because I had a fair bit of responsibility and very little power.
I'd had much more ability to control things at UCL.
Once I got to Cambridge,
and I was submerged in the larger mathematics department,
I discovered that I didn't have nearly as much power to control things
as I thought I would have.
So that was very frustrating.

\textbf{Vovk:}
In 1992 you were Joint Editor of \emph{JRSS B},
and also you were Editor of \emph{Biometrika} for four years,
and then of \emph{Bayesian Analysis} for four years.
Did it take much of your time?
And was it useful for your own research?
Maybe it broadened your horizons\dots.

\textbf{Dawid:}
Just to clarify, I was, yes, editor of \emph{Series B},
then I got offered the \emph{Biometrika} gig,
so I gave up \emph{Series B}.
As for \emph{Bayesian Analysis}, I was never the top editor.
They may have called us editors, but it was basically Associate Editor.

The answer to the first question is yes, I think.
\emph{Biometrika} in particular was pretty time consuming,
because there was a high submission rate,
and quite a lot of work for the editor to do.
Did it stimulate my own research?
There may have been one or two papers that passed by
which gave me  some ideas,
but I think the answer is probably not a lot.
For \emph{Biometrika}, the kind of topics which it dealt with
were interesting and important, but often a little bit far from my own core interests.

\textbf{Vovk:}
Your work has been widely recognized.
You are a Fellow of the Royal Society.
What kind of position is it?
Is it an honorary position,
or does it involve a lot of work, like awarding grants or promoting their goals in different ways?

\textbf{Dawid:}
The main job of the Royal Society is self-preservation.
I remember the speech that the President made to the new fellows;
he said that the society is 85\% gonad.
It's all about reproducing itself.
So it's basically about suggesting new fellows and vetting them.
I've served for three years on what they call the sectional committee for Mathematics,
which is all about making proposals for new fellows in that area.
Of course, it goes to a higher committee for final decision.
In terms of what I was asked to do, yes, that was the main thing.
There wasn't much else that came my way,
although they do have lots of other activities, obviously.

\textbf{Vovk:}
In 2000 you were President of the International Society for Bayesian Analysis.
What was your role as President?

\textbf{Dawid:}
To be President was largely a ceremonial role at that time.
I had to make a few speeches,
but, for example, I didn't have a lot to do with the actual organization of their meetings.

\textbf{Shafer:}
What was the importance of having a separate Bayesian society?
Why wouldn't the existing societies be adequate for your purposes?

\textbf{Dawid:}
This is interesting.
In the early days there was quite a lot of discussion
about whether it was indeed appropriate to have a separate society,
and a similar discussion about whether it would be appropriate
to have a specialist Bayesian journal,
which eventually came out.
A lot of people said no, no, no,
you should just do your Bayesian work within the standard societies and journals
and spread it around,
and it will be dangerous to hive it off,
where it could be more easily put to the side.
If you put it in a container, you can drop the container overboard,
and then we'll never think about it again.
So there was a lot of concern, a lot of worries about it.
Both the society and the journal; were they good ideas or not?
And I shared some of the misgivings in the early days,
but there's no doubt that they were good ideas, as it turned out,
because the sheer volume of Bayesian stuff needed outlets.
A lot of it is still coming out in regular journals and all over the place anyway,
but it is a good thing to have specialist places where it will be welcome.
We know it will be welcome both in the conferences and in the journal.
And I think the very fact that those, the society and the journal, exist
has stimulated a lot of further development
and what has now become a completely amazing expansion of Bayesian statistics into the big wide world.
When I was a lad and there were only 30 Bayesians in the world,
we couldn't ever conceive it.

\section{Other interests}

\textbf{Vovk:}
What do you enjoy doing when you are not doing statistics?

\textbf{Dawid:}
I'm pretty lazy, and I don't have a lot of outside interests really.
Nothing of any great excitement.
I enjoy listening to music and going for walks.
Seeing the family,
all the boring, straightforward, standard things.
I don't have any exciting hobbies.
After I retired, having tried and failed to play the piano when I was young,
I thought I'd try again, which I did, and I failed again [laughs].
So I realized I didn't have what it takes,
but luckily, my grandchildren make up for it because they are all splendid musicians.
So it's in the family.

\textbf{Vovk:}
When did you meet your wife, Elah\'e, and what's her profession?

\textbf{Dawid:}
She was a trained as a doctor,
and she'd gone back to Iran at some point,
hoping to make a career there,
but realized it wasn't going to work out.
She became interested in family planning.
Anyway, she came back to to England,
and there was a course on medical demography at the London School of Hygiene and Tropical Medicine.
She enrolled on that.
She was still, I think, planning to go back to Iran at that point.
So she was at the London School of Hygiene,
which at that time was the venue where the Royal Statistical Society would hold its discussion meetings.
At one point there was a rather famous meeting by Mervyn Stone
on cross-validatory choice and assessment of statistical procedures.
And he gave that talk, and I'd prepared a discussion of the talk,
and I got up, gave my discussion, and came and sat down again.
And then the lady in the row behind tapped me on the back,
and basically said,
where are you from?
I said University College,
but I think she meant what was my ethnic background [smiles].
Anyway that was my introduction to Elah\'e.

\textbf{Vovk:}
What about your children?
Are they interested in science?

\textbf{Dawid:}
No, not really.
Neither my son Jonathan nor my daughter Julie.
Jonathan studied physical sciences at Cambridge
and then spent a year at Harvard to do a PhD.
But he lost interest, did a postgraduate conversion course in law,
and for many years has been a barrister.

\textbf{Vovk:}
What are your views on higher education in the UK?
For example, what do you think about the REF (Research Excellence Framework) exercise?

\textbf{Dawid:}
Well, I'm out of it now, thank God.
It was frightful at the time.
Did it improve higher education?
I rather doubt.
It introduced a tremendous lot of bureaucratic hassle, an amazing amount of hassle.
It introduced a sort of market in individuals.
Departments would raise their profile not by doing better,
but by buying in stars.
Which is a bit more like football teams than university departments.
I always felt that my colleagues and I knew what we should be doing,
the purpose was to do it well,
and to have people looking over our shoulder and assessing us all the time
was not the way to do things.
So I don't really approve of those massive disruptive exercises
which only distort the thing that they're trying to measure.

\section{Future of statistics and data analysis}

\textbf{Vovk:}
How do you see statistics as a component in a bigger field,
which probably might be called data analysis?
I know that some statisticians, such as George Box,
consider data analysis to be part of statistics,
but I mean data analysis as the union of different communities
such as machine learning, statistics, data mining, etc.

\textbf{Dawid:}
Yes, they are different communities,
which is a pity really, because there's a lot of common ground.
It's increasingly realized how much common ground there is.
And there's more and more flow of people and ideas between these different subjects.
Essentially they could be regarded
as different but closely related corners of a larger enterprise,
so a lot of people who train as statisticians will end up in jobs called data scientists
or be coming into machine learning departments, or whatever.
And so much good work in statistics is being done out in those otherwise named departments.
I don't care what people call themselves, as long as they're doing good work, and a lot of it is being done.
That's lovely.
There was a time early on when I think there was a lot of rediscovering the wheel
as new groups that came into being gave themselves new names and didn't really know a lot of what had gone before.
But now, I think, there's a lot of appreciation.
There are obviously different emphases:
whether you emphasize logic, whether you emphasize computation, whether you emphasize applications.
But it's great that there is this
broadening of the whole scope of statistical science.
So I am all in favour.

\textbf{Vovk:}
A somewhat related question is about the role of statistics in science.
It's been fascinating for many statisticians,
and it was a major concern, for example, for Fisher and George Box.
What do you think about it?
Things like the recent talk of reproducibility,
or p-values versus Bayes factors.

\textbf{Dawid:}
You can say that the way statistics has been taught over the years
has had some unexpected dangerous side effects.
The emphasis on significance testing, for example, or accept/reject,
rightly or wrongly is being perceived by a lot of scientists
as ``is this paper publishable or not publishable?''\ rather than
``have I discovered something worth knowing or not worth knowing?'',
and knowing that a hypothesis that you previously hoped was true is false
is of course something worth knowing.
The whole gatekeeping role of statistics led
to a complete distortion of what got published in the scientific journals.
Reproducibility is one aspect of that,
but the essential problem is the selection effect:
what gets to be known about is highly uncorrelated with what is worth knowing about.
So that has been a problem, and I'm not quite sure how to solve it.
Just bringing alternatives to p-values,
even e-values, how can you stop them being misused?
It's not clear to me that you can.
Whatever you do, however clever the idea and however worthy, and however much you try and explain.
There's a sense in which statisticians were gatekeepers.
We wanted to separate out good work from bad work.
But that has been misunderstood as separating out significant from insignificant [smiles],
or variations on that using other approaches to statistics.
So how do we tell what's good and what's bad?
That's what statistics should be about.
Sound data and sound science.

\textbf{Vovk:}
These are serious problems faced by any approach to testing.
But let's assume for a minute that we are talking about analysing
results of a well-planned and well-executed experiment.
Could you summarize your views of p-values in science?
Do you think p-values per se are harmful in science,
or is it a danger just for inexperienced scientists?
Perhaps scientists need to be educated.

\textbf{Dawid:}
We can think of p-values
in terms of the simple logic of Fisherian significance testing.
There's a link to something we talked about earlier, Cournot's principle.
If you have a hypothesis,
and you observe something that is simple to describe
but is improbable under that hypothesis,
then maybe you should rethink your hypothesis.
And there is a grain of value in that, no doubt about it.
What's the best way to formalize it, is another matter.
So I say yes, there's a grain of value
in p-values, e-values, whatever you want.
And of course, Cournot's principle is fundamental to interpretations of game-theoretic probability,
as I don't have to tell you.

\textbf{Vovk:}
What about Bayes factors?
Is it something that you accept?

\textbf{Dawid:}
To deviate very slightly,
the forensic statistics section tried to spread the light
and emphasize the role of likelihood ratios in forensic reasoning,
considering that usually there's a hypothesis for the prosecution
and a hypothesis for the defence.
As statisticians, we express that probabilistically,
then we think the way to decide between them is to look at the likelihood ratio,
Bayes factor if you like.
And so a lot of what forensic statisticians have been trying to do
is to persuade people in the legal and forensic professions
that the right way to think about weighing up the two arguments is through likelihood ratio.
I think that is generally an excellent idea.
So I'm in favour of Bayes factors,
although personally, I prefer posterior odds.
But in the legal context you're not allowed to do that.
What you can sensibly talk about and hope to be able to transmit with agreed understanding,
which isn't easy,
is likelihood ratios and Bayes factors.
So yes, I'm all in favour.
Indeed, likelihood functions more generally;
it doesn't have to be just two hypotheses.
Sometimes you have more, which people seem to forget.

\textbf{Vovk:}
In conclusion,
what would be your advice for young statisticians,
theoretical and applied?

\textbf{Dawid:}
Different people will have different interests, naturally.
What's the most important thing?
Let me just make a confession.
I am not really keeping up with modern theoretical statistics [laughs].
Only very, very small corners of it which tickle my own interests.
There's a lot of stuff out there which is extremely important.

It's been interesting to see the change, for example,
in the Bayesian camp.
While in the early days the emphasis was, for lack of computation,
almost entirely on theory, methodology, and logic,
now it's almost entirely on computation and applications,
and it's wonderful and incredibly inspiring.
I think the big action is in applications these days.
Genetics is obviously one of the biggest.

\textbf{Shafer:}
Thank you, Philip, for this informative and inspiring conversation.

\subsection*{Acknowledgments}

\noindent
We are grateful to Sonia Petrone,
the Editor of \emph{Statistical Science} at the time,
and to Pietro Rigo and Bertrand Clark,
the Guest Editors of the Special Issue of \emph{Statistical Science}
on prediction,
for the idea of this conversation.

\end{document}